\documentclass[10pt,journal,compsoc]{IEEEtran}
\ifCLASSOPTIONcompsoc
  \usepackage[nocompress]{cite}
\else
  \usepackage{cite}
\fi
\ifCLASSINFOpdf
\else
\fi
\usepackage[linesnumbered,noline,boxed,commentsnumbered, ruled]{algorithm2e}
\usepackage{float}
\usepackage{booktabs}
\usepackage{flushend}
\usepackage{multirow}
\usepackage{mathrsfs}
\usepackage{lscape}
\usepackage{bm}
\usepackage{tikz,pgf}
\usepackage{doi}
\usetikzlibrary{arrows,shapes,positioning,shadows,trees}

\newcommand{\reffig}[1]{Figure \ref{#1}}
\newcommand{\refsec}[1]{Section \ref{#1}}
\newcommand{\refequ}[1]{Equation (\ref{#1})}
\usepackage{enumerate}
\usepackage{subfigure,graphicx,color}
\usepackage{amsfonts,amsmath,amssymb}

\begin{document}
%
\title{Improving Dither Modulation based Robust Steganography by Overflow Suppression}
%
%
%
%

\author{\IEEEauthorblockN{Kai Zeng, 
Kejiang Chen, 
Weiming Zhang, 
Yaofei Wang, 
Nenghai Yu}
\IEEEcompsocitemizethanks{\IEEEcompsocthanksitem This work was supported in part by the Natural Science Foundation of China under Grant 62002334 and 62072421, and by Anhui Science Foundation of China under Grant 2008085QF296, and by Anhui Initiative in Quantum Information Technologies under Grant AHY150400.
\IEEEcompsocthanksitem All the authors are with Key Laboratory of Electromagnetic Space Information, School of Information Science and Technology, University of Science and Technology of China, Hefei 230026, China.
\IEEEcompsocthanksitem Corresponding authors: Kejiang Chen and Weiming Zhang (Email:chenkj@ustc.edu.cn, zhangwm@ustc.edu.cn).}}

\IEEEtitleabstractindextext{%
\begin{abstract} Nowadays, people are sharing their pictures on online social networks (OSNs), so OSN is a good platform for Steganography. But OSNs usually perform JPEG compression on the uploaded image, which will invalidate most of the existing steganography algorithms. Recently, some works try to design robust steganography which can resist JPEG compression, such as Dither Modulation-based robust Adaptive Steganography (DMAS) and Generalized dither Modulation-based robust Adaptive Steganography (GMAS). They relieve the problem that the receivers cannot extract the message correctly when the quality factor of channel JPEG compression is larger than that of cover images. However, they only can realize limited resistance to detection and compression due to robust domain selection. To overcome this problem, we meticulously explore three lossy operations in the JPEG recompression and discover that the key problem is spatial overflow. Then two preprocessing methods Overall Scaling (OS) and Specific Truncation (ST) are presented to remove overflow before message embedding as well as generate a reference image. The reference image is employed as the guidance to build asymmetric distortion for removing overflow during embedding. Experimental results show that the proposed methods significantly surpass GMAS in terms of security and achieve comparable robustness.
\end{abstract}

}

\maketitle

\IEEEdisplaynontitleabstractindextext

%
\IEEEpeerreviewmaketitle

\IEEEraisesectionheading{\section{Introduction}}

%
%
%
%

\IEEEPARstart{S}{teganography} is a science and art of covert communication that transmits secret messages through digital media without attracting the attention of others~\cite{IHTDSC,SteTDSC}. Many different mediums can be used in steganography, such as texts, audios, images and videos. Among them, JPEG images are now widely used in people's lives because they can provide high-level visual quality with low storage costs~\cite{JPEG}. At present, the most remarkable steganographic schemes at JPEG images are based on the framework of minimizing distortion~\cite{JPEGSteTDSC,JPEGSteTDSC2}. Within this framework, researchers define a modification cost to each cover element and use Syndrome-Trellis Codes (STCs)~\cite{STCs} to embed message, where STCs can asymptotically reach the payload-distortion bound for additive distortion. 

With the practical steganographic code STCs, the latest researches focused on how to design effective distortion function. There are a lot of distortion functions about JPEG image, such as J-UNIWARD (JPEG UNIverlet WAvelet Relative Distortion)~\cite{JUNIWARD}, UERD (Uniform Embedding Revisited Distortion)~\cite{UERD}, RBV (Residual Block Value)~\cite{RBV}, BET (Block Entropy Transformation)~\cite{BET}, GUED (Generalized Uniform Embedding Distortion)~\cite{GUED} and J-MiPOD (Minimizing the Power of Optimal Detector for JPEG domain)~\cite{J-MiPOD}. Their main purpose is to assign low costs to the coefficients in complex areas of an image and high costs to the coefficients in smooth areas according to the “Complexity-First Rule”~\cite{Complexity-First-Rule}. Non-additive cost functions on JPEG images are defined to keep the continuity of adjacent blocks in the spatial domain~\cite{BBC,BBC++,BBM}.


The aforementioned works have been performed well on lossless channels. However, images transmitted over online social networks (OSNs), such as Facebook and Twitter, generally suffer from lossy processes. To ensure image quality while saving storage space, OSN usually performs JPEG compression on the uploaded images. JPEG recompression will invalidate the traditional steganography scheme because the stego changes during transmission~\cite{DSTC}. Therefore, JPEG recompression resistance is the major concern due to the wide use of the JPEG recompression in OSNs~\cite{SNP}. In recent years, there has been a growing number of publications~\cite{TCM,DMAS,Robust-steganography-framework,GMAS,Downward-robust,AE-RS} focusing on the steganography in this real-world communication and aiming to resist these lossy processes. 

We summarize existing steganographic algorithms against JPEG compression into two categories according to their application scenarios. The first category performs well only if we know the quality factor (QF) for channel recompression or can use the channel at will, and is called ``White-Box Robust''. The representative of this category is TCM (Transport Channel Matching) proposed by Zhao \textit{et al.}~\cite{TCM}, which repeatedly processes the image by applying channel manipulations until the image is nearly identical before and after processing. They additionally utilized the Error Correction Code (ECC) to reduce the Bit Error Rate (BER). However, this method requires repeatedly uploading and downloading cover images on OSNs, which is abnormal behaviour and will arouse the suspicion of the attacker.

The other category, which does not require a priori knowledge of the channel and cannot be used arbitrarily, is named ``Black-Box Robust''. Zhang \textit{et al.}~\cite{DMAS} proposed Dither Modulation-based robust Adaptive Steganography (DMAS) based on ``Robust Domain Selection + ECC-STCs Codes''~\cite{Robust-steganography-framework} to achieve undetectability and robustness in this situation. This approach avoids behavioral insecurity. They chose middle frequency AC (Alternating Current) coefficients as the cover and utilized dither modulation for message embedding. Yu \textit{et al.}~\cite{GMAS} further upgraded dither modulation to generalized dither modulation by expanding robust domain and introducing asymmetric distortion in Generalized dither Modulation-based robust Adaptive Steganography (GMAS). However, they only perform well when the QF of the cover image is less than that of the channel recompression, which is called ``Upward Robust''. Concerning the other case that the QF of channel JPEG compression is less than that of cover image, Tao \emph{et al.}~\cite{Downward-robust} proposed an enjoyable scheme. They deduce the JPEG compression process in mathematical to adjust the DCT coefficients so that the receiver can exactly extract the secret messages from the stego after compression. Nevertheless, this method does not consider the rounding and truncation in the spatial domain so is hardly used in practice. Thus, designing the algorithm available in this case is still an open problem.

In summary, black-box robust steganography, such as GMAS and DMAS, are the current practical algorithms with wide application scope, yet they can realize desirable resistance to detection and compression only at low capacity due to the limitations of robust domain selection. To overcome these problems, we first explore the reasons for the errors of previous upward robust steganography algorithms: lossy operations in the JPEG recompression. Then we investigate the impacts caused by spatial operations, a key issue that has not received much attention in previous studies, and find that spatial overflow is the severest operation. Based on this finding, we propose a novel robust steganographic method by cover-preprocessing and asymmetric cost definition. In detail, cover-preprocessing can effectively reduce spatial overflow to enhance the robustness of the coefficient as well as generate a reference image. The asymmetric distortion is designed according to the reference images to make steganography towards the direction which produces a more robust image. The security of the proposed scheme is verified with detailed experiments under different compression QFs and valid steganalysis with CCPEV (PEV features~\cite{CC} enhanced by Cartesian Calibration)~\cite{CCPEV}, DCTR (Discrete Cosine Transform Residual)~\cite{DCTR} and SRNet~\cite{SRNet}. Notably, we use the image before preprocessing as cover and image through steganography as stego for steganalysis when evaluating the security of the algorithm, which is more practical than the assessment method used in~\cite{TCM}. The experimental results show that the proposed scheme can reach higher robustness and security compared to previous algorithms.

The contributions of this work are summarized as follows.
\begin{enumerate}[1)]
\item This paper explores the lossy process of JPEG compression in detail and finds that the key problem in anti-compression steganography is spatial overflow. Reducing the impact of spatial truncation operations is a effective way to improve robust steganography.

\item With the analysis of the JPEG compression, we first propose a preprocessing method that completely removes spatial overflow and theoretically proves its effectiveness. In the pursuit of greater security, a heuristic preprocessing method with fewer modifications is proposed. Both methods enhance the stability of the coefficients,  allowing the entire image to be used as a robust region.

\item Considering both security and robustness, a novel asymmetric distortion definition method is proposed, which combines the embedding process with the removal of the spatial overflow together.

\end{enumerate}

The rest of this paper is organized as follows. We introduce the related work in~\refsec{section2}. An investigation of the JPEG recompression on the dither modulation based algorithms is in \refsec{section3}. The exhaustive process of the proposed scheme are described in ~\refsec{section4}. The consequences of contrast experiments and tests on their performance are shown in~\refsec{section5}. Finally, ~\refsec{section6} concludes the paper.

\section{Related Work}\label{section2}

\subsection{Notations and JPEG Recompression}\label{notations}

Throughout the paper, matrices, vectors and sets are written in capital letters. Unless otherwise indicated, we use symbol $\boldsymbol{D}$ to denote the matrices of quantized DCT coefficients, $\boldsymbol{Q}$ to denote the matrices of quantization table, $\boldsymbol{S}$ to denote the matrices of spatial values derived from $\boldsymbol{D}$ by IDCT and $\boldsymbol{X}$ to denote the matrices of pixels. Elements in the matrix are denoted by corresponding lowercase letters, for example $\boldsymbol{D}=(d_{u,v})$, $\boldsymbol{Q}=(q_{u,v})$, $\boldsymbol{S}=(s_{i,j})$, $\boldsymbol{X}=(x_{i,j})$, where the subscripts mean their position. Considering the way JPEG compression chunking operates we normally use $8\times8$ block, namely, $u,v,i,j\in\{1,2,\cdots,8\}$. Also considering the storage limitations in JPEG, $d_{u,v}\;(or\;d)\in\{-1024,\cdots ,1023\}$, $q_{u,v}\;(or\;q)\in\{1,\cdots ,255\}$, $s_{i,j}\;(or\;s)\in\{-128,\cdots,127\}$ and $x_{i,j}\;(or\;x)\in\{0,\cdots,255\}$. The transformation functions are denoted by bold symbols, additionally, $\left[\;\cdot\;\right]$, $\left\lceil\;\cdot\;\right\rceil$ and $\left\lfloor\;\cdot\;\right\rfloor$ are denote rounding methods round, ceil and floor in mathematics. The aforesaid notations and the relationship among the above elements can be briefly explained through the JPEG compression process as follows.

The JPEG recompression process first converts the DCT coefficients to spatial values, then truncates and rounds them to obtain spatial image, finally converts spatial image into quantized DCT coefficients with the QF of channel. For detail, the process of recompressing an $8\times8$ DCT block $\boldsymbol{D}$ from quality factor $q_1$ to $q_2$ is as follows:
\begin{itemize}
\item\emph{coefficients dequantization:}
\begin{equation}\label{REa}
\boldsymbol{\widetilde D}_{q_1}=\boldsymbol{D}_{q_1}\times \boldsymbol{Q}_{q_1}, 
\end{equation}
\item\emph{inverse discrete cosine transformation:}
\begin{equation}\label{REb}
\boldsymbol{S}=\textbf{IDCT}(\boldsymbol{\widetilde D}),
\end{equation}
\item \textbf{\emph{spatial truncation:}}
\begin{equation}\label{REc}
\boldsymbol{\dot S}=\textbf{TRU}(\boldsymbol{S}),
\end{equation}
\item \textbf{\emph{spatial rounding:}}
\begin{equation}\label{REd}
\boldsymbol{X}=\left[\boldsymbol{\dot S}+128\right],
\end{equation}
\item\emph{spatial shift:}
\begin{equation}\label{REe}
\boldsymbol{\ddot S}=\boldsymbol{X}-128,
\end{equation}
\item\emph{discrete cosine transformation:}
\begin{equation}\label{REf}
\boldsymbol{\widetilde D}_{q_2}=\textbf{DCT}(\boldsymbol{\ddot S}),
\end{equation}
\item \textbf{\emph{coefficients quantization:}}
\begin{equation}\label{REg}
\boldsymbol{D}_{q_2}=\left[\boldsymbol{\widetilde D}_{q_2}/ \boldsymbol{Q}_{q_2}\right].
\end{equation}
\end{itemize}
We use $\boldsymbol{\widetilde D}$ to denote the matrices of unquantized DCT coefficients. The function \textbf{TRU}$(\;\cdot\;)$ is truncation of the pixel which value is out of the spatial range thus called overflow:
\begin{equation}\label{truncation}
\textbf{TRU}(x)=\left\{\begin{array}{lc}127&x>127\\-128&x<-128\\x&else\end{array}\right..
\end{equation}

Notably, DCT coefficient quantization referred to in this paper usually includes both quantization and rounding operations unless specified. Greek letters denote costs of the coefficients and parameters. Specific explanations of the notations, or elements not mentioned, will be elaborated on later.

\subsection{Dither Modulation}\label{subsection2.1}

\begin{figure}[t]
\centering
\includegraphics[width=3.3in]{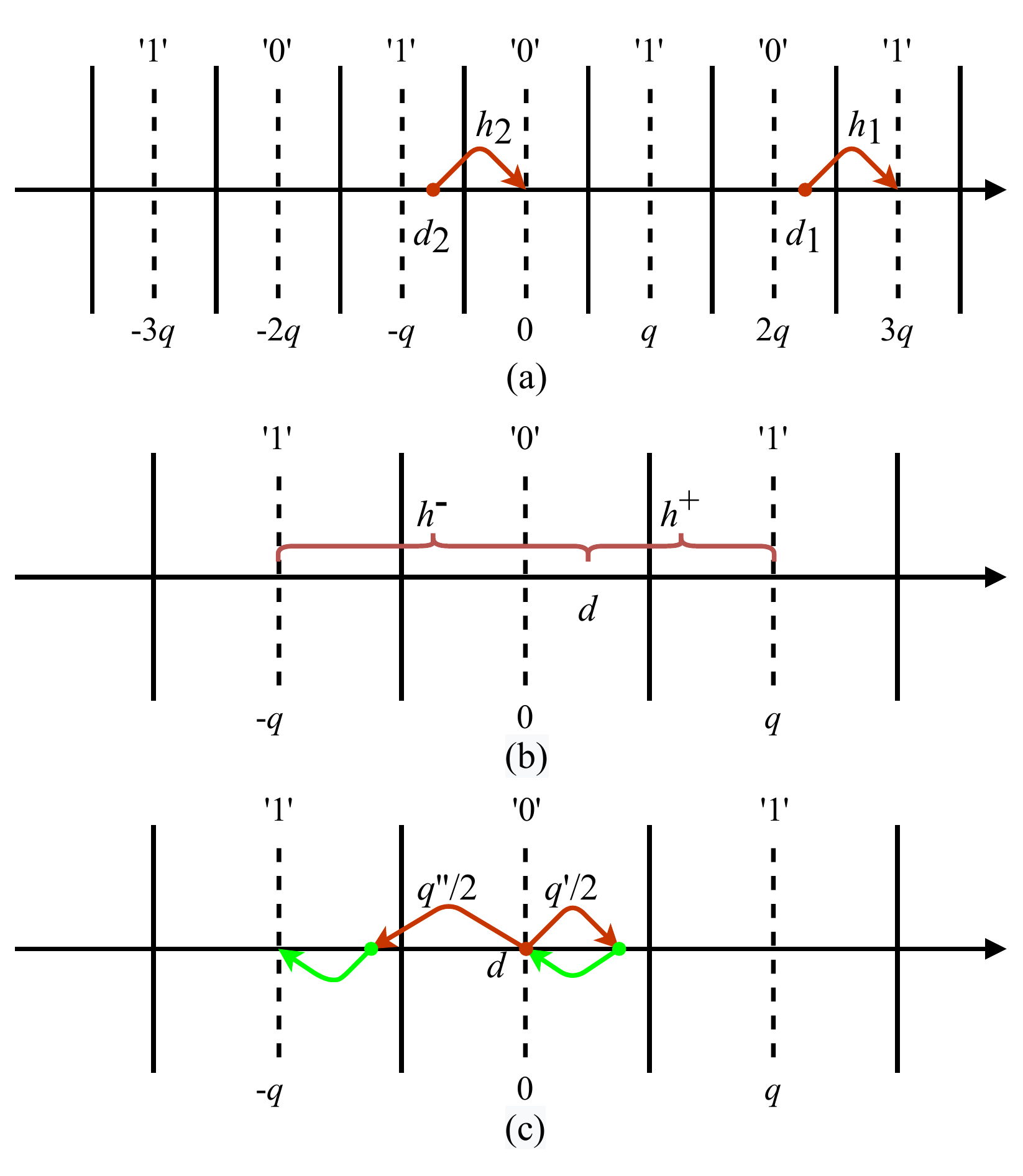}
\caption{\small{Embedding scheme in different circumstances : (a) Dither modulation in the watermarking algorithm and DMAS; (b) Generalized dither modulation in GMAS; (c) The situation discussed in \refsec{subsection3.1} .}}
\label{fig-DM}
\end{figure}

Dither modulation is an extension of the original uniform quantized index modulation (QIM) algorithm~\cite{QIM}. Since dither modulation can be robust in embedding while maintaining high visual quality of the image, it is widely used in watermarking algorithms. In the below, we briefly introduce the embedding process of dither modification on DCT coefficients in~\cite{QIM-in-DCT}.

The modification method of algorithm~\cite{QIM-in-DCT} is shown in \reffig{fig-DM}(a). We use $d_1$ and $d_2$ to represent the unquantized DCT coefficients and the quantization step associated with them is $q$. The axes which represent the unquantized DCT coefficients can be divided into many segments by quantization step $q$. As seen in the \reffig{fig-DM}(a), $d_1$ in the odd segment represents the message bit ‘1’, and $d_2$ in the even segment represents the message bit ‘0’. If we want $d_2$ to represent the message bit ‘0’, we will add $h_2$ to it and adjust it to the center of the nearest even segment. A similar modification will be performed if the message represented by $d_1$ is changed.

\subsection{DMAS and GMAS}\label{subsection2.2}

Dither Modulation-based robust Adaptive Steganography (DMAS)~\cite{DMAS} is a recent robust steganographic algorithm under the “Robust Domain Selection + ECC-STCs Codes” framework. It constructs compression-resistant domain mainly by choosing middle frequency DCT coefficients as the embedding domain in conjunction with dither modulation. DMAS can be applied in either spatial or DCT domain. 

Generalized dither Modulation-based robust Adaptive Steganography (GMAS)~\cite{GMAS} improves resistance to detection and compression by generalized dither modulation and expanding the embedding domain of DMAS. The embedding scheme of generalized dither modulation is illustrated in the \reffig{fig-DM}(b). Unlike DMAS, which modifies the coefficients to the center of the nearest interval. GMAS considers the possibility of modifying to both sides as well as introduces ternary embedding and asymmetric distortion for hiding message. We will only describe GMAS in detail here:
\begin{enumerate}[i.]
\item Select Cover Elements from Expended Robust Domain. The expended embedding domain are composed of coefficients $d_{i,j}$ satisfying the $(i+j=7,8,9)$. These coefficients are selected as steganographic cover and embedded using generalized dither modulation.
\item Calculate Asymmetric Distortion. Calculate the spatial pixels that eliminate the JPEG compression block effect. 
\begin{equation}\label{equation4}
\boldsymbol{X'}=\boldsymbol{X} \otimes \boldsymbol{F},
\end{equation}
\begin{equation}\label{equation5}
\boldsymbol{F}=\begin{bmatrix}1/9&1/9&1/9\\1/9&1/9&1/9\\1/9&1/9&1/9\end{bmatrix}.
\end{equation}
To improve the performance, use this $\boldsymbol{X'}$ as a reference and define asymmetric distortion to encourage modifications toward the reference image. In detail, a block DCT transformation of $\boldsymbol{X'}$ yields unquantized DCT coefficients ${\overline d}_{i,j}$, calculate symmetric costs $\rho_{i,j}$ with existing distortion functions then asymmetric costs of the quantized DCT coefficients can be defined as follows:
\begin{equation}\label{equation6}
\rho_{i,j}^+=\left\{\begin{array}{l}\lambda\cdot\rho_{i,j}\;,\;d_{i,j}<{\overline d}_{i,j}/q_{i,j}\\\rho_{i,j}\;,\;else\end{array}\right.,
\end{equation}
\begin{equation}\label{equation7}
\rho_{i,j}^-=\left\{\begin{array}{l}\lambda\cdot\rho_{i,j}\;,\;d_{i,j}>{\overline d}_{i,j}/q_{i,j}\\\rho_{i,j}\;,\;else\end{array}\right.,
\end{equation}
where $d_{i,j}$ denotes the quantized DCT coefficients and $\lambda\in(0,1)$ controls the intensity of the adjustment. The final asymmetric costs $\xi_{i,j}^+$ and $\xi_{i,j}^-$ for the unquantized DCT coefficients are calculated as follow:
\begin{equation}\label{equation8}
\begin{aligned}
\zeta_{i,j}^+&=\rho_{i,j}^+/q_{i,j},\\h_{i,j}^+&=(k+1) q_{i,j}-\tilde d_{i,j}, \\ \xi_{i,j}^+&=\zeta_{i,j}^+\times h_{i,j}^+,
\end{aligned}
\end{equation}
\begin{equation}\label{equation9}
\begin{aligned}
\zeta_{i,j}^-&=\rho_{i,j}^-/ q_{i,j},\\h_{i,j}^-&=\tilde d_{i,j}-(k-1)q_{i,j}, \\ \xi_{i,j}^-&=\zeta_{i,j}^-\times h_{i,j}^-,
\end{aligned}
\end{equation}
where $k\in N$ is an integer. $\tilde d_{i,j}\in((k-1/2)q_{i,j},(k+1/2)q_{i,j})$ and $q_{i,j}$ denote the unquantized DCT coefficients and the corresponding quantization step, respectively.
\item RS Encoding and Ternary STC Embedding. Utilizing RS codes to encode the messages. A modifiable ternary sequence can be obtained from the selected coefficients using generalized dither modulation so ternary STC is used to embed to attain a lower bit error rate and stronger security. Specifically, ternary STC can be implemented using double-layered STCs~\cite{DBL}.
\end{enumerate}

After receiving the compressed stego image, the receiver first converts JPEG images to spatial values, then obtains the unquantized DCT coefficients derived from spatial images, finally calculates the quantized DCT coefficients with the same quantization table as that used for embedding, and we call this entire operation “\textbf{DCT Coefficients Restoration}”. After that, the same middle frequency DCT coefficients are chosen to gain the message using STCs and RS decoding.

\section{Anti-compression Analysis of Dither Modulation}\label{section3}

\begin{figure*}[t]
\hspace{-5mm}\centering\subfigure[Original Block]{
\includegraphics[width=1.8in]{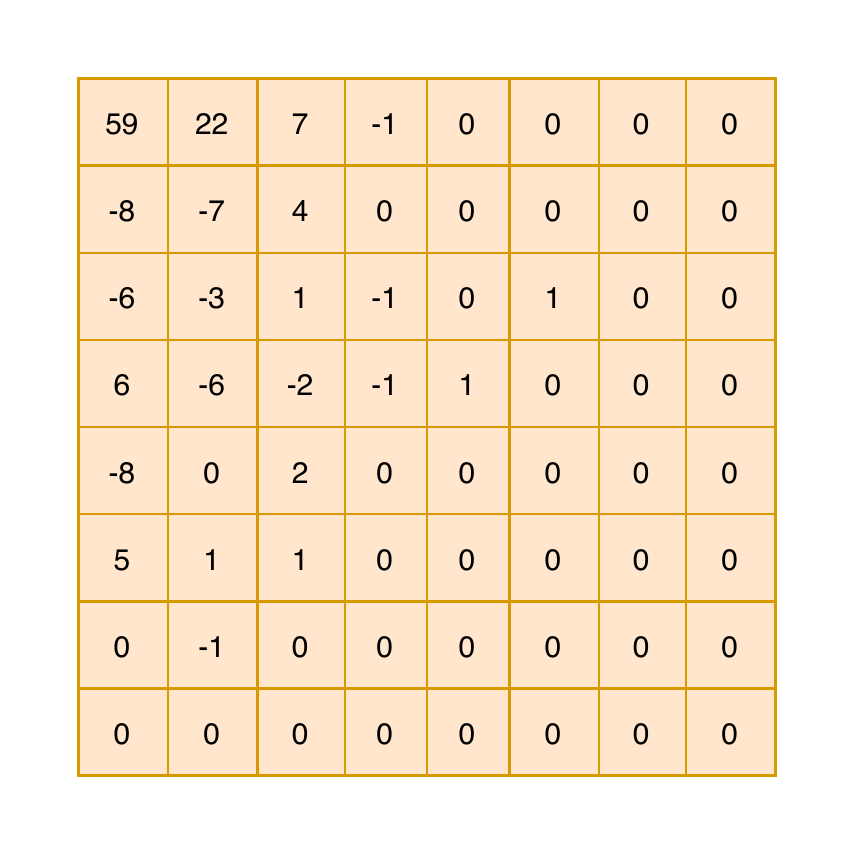}}\hspace{-5mm} \centering\subfigure[Truncation Loss]{
\includegraphics[width=2.8in]{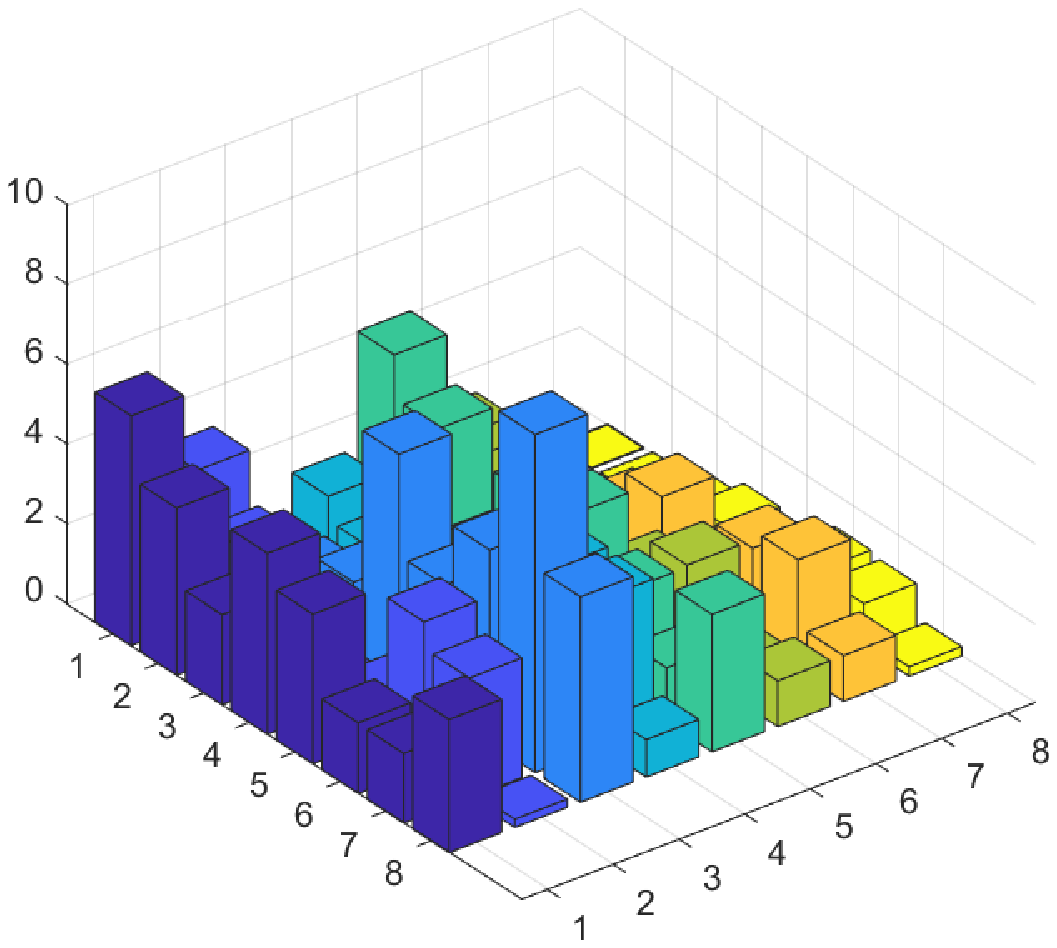}}\hspace{-10mm}
\centering\subfigure[Rounding Loss]{
\includegraphics[width=2.8in]{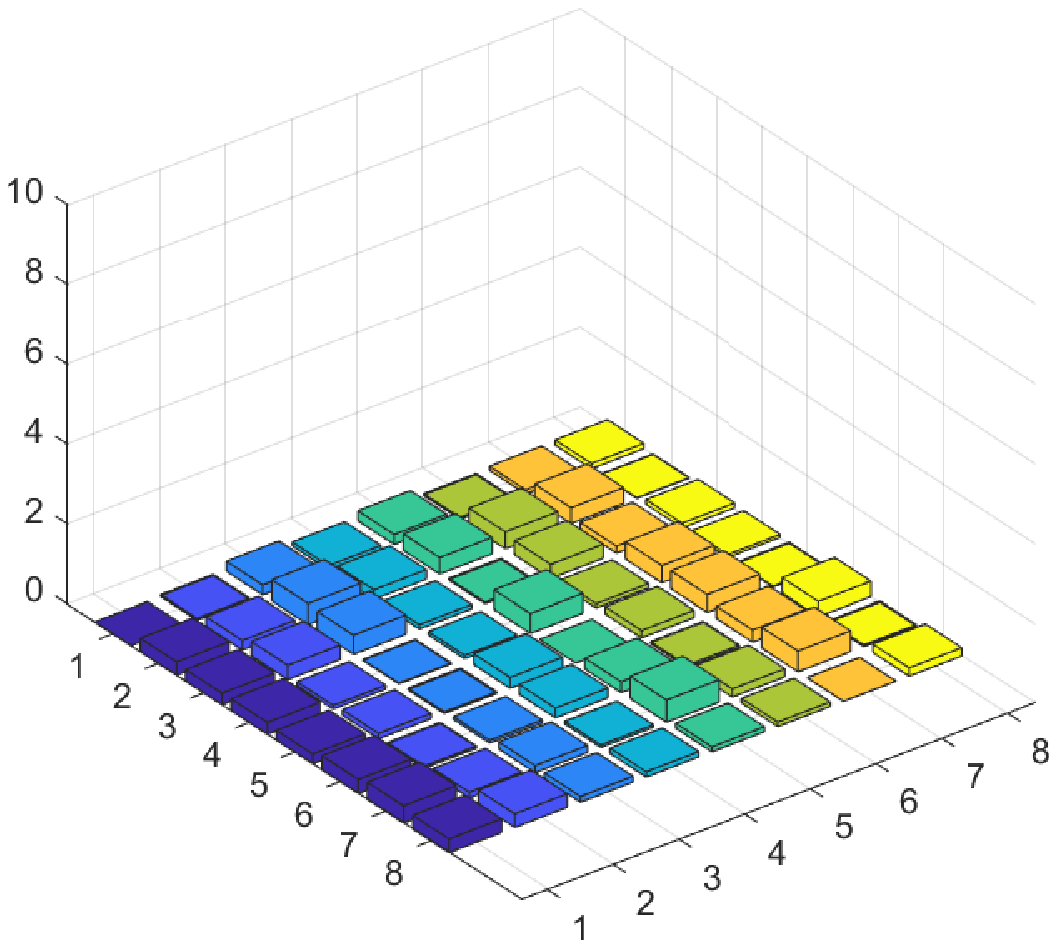}}\hspace{-10mm}
\caption{\small{Effect of spatial truncation and rounding operations on specific DCT coefficients. Randomly select a block from an image in BOSSbase 1.01~\cite{Bossbase} with QF = 65. (a) demonstrates the value of the concrete quantized DCT coefficients. We then convert this block to the spatial values and perform respectively truncation and rounding operations at each position. (b) and (c) illustrate the absolute values of the modifications on the regained unquantized DCT coefficients compared to the original.}}
\label{fig-TQ}
\end{figure*}

State-of-the-art black-box robust steganography is based on dither modulation, which is the main reason why they do not require knowledge of channel to achieve robustness. In this section we will explore the advantages and disadvantages of this approach. 

As displayed in \refsec{notations}, there are three lossy steps in the recompression process: \emph{spatial truncation}, \emph{spatial rounding} and \emph{coefficients quantization}. These three operations have different effects on the DCT coefficients. Let's start our detailed analysis by probing the principle of upward robust.

\subsection{Explanation of Upward Robustness}\label{subsection3.1}

Yu \textit{et al.}~\cite{GMAS} experimentally found that steganographic algorithms using dither modulation for embedding (such as DMAS and GMAS) generally behave excellently when the QF of channel compression is larger than that of the cover image and called this phenomenon ``Upward Robust''. This part will provide a theoretical analysis of this phenomenon. 

Generally, we denote the unquantized DCT coefficient of a stego JPEG image by $\tilde d$. Then the relation among $\tilde d$ , the corresponding quantized coefficients $d$, and quantization steps $q$ is as follows:
\begin{equation}\label{equation10}
\tilde d=d\cdot q.
\end{equation}
This means that the unquantized DCT coefficients of the JPEG image are exactly at the center of the region divided by the quantization step. If the quantization step used for this coefficient in channel compression is $q'$, the maximum modifications caused by channel compression is $q'/2$ without considering the processing of the spatial pixels. So after channel compression we have:
\begin{equation}\label{equation11}
\tilde d_{q'}\in\lbrack \tilde d-\frac{q'}2,\tilde d+\frac{q'}2\rbrack,
\end{equation}
where $\tilde d_{q'}$ is unquantized DCT coefficient after channel compression. As described in the \refsec{subsection2.2}, it is necessary to perform DCT coefficients restoration. Denote the coefficient after restoration at this situation as $d_{q'}$. When the channel compression QF is larger than the cover QF (the corresponding quantization step of channel compression is less than that of the cover), it is obvious from the \reffig{fig-DM}(c) that
\begin{equation}\label{equation12}
\frac {q'}2<\frac q2\;,\;\tilde d_{q'}\;mod\;q\;=\;\tilde d\;mod\;q\;,\;d\;=\;d_{q'}.
\end{equation}
In this case, the coefficients after coefficients restoration are the same as the original ones, so the message can be extracted accurately. If the QF of channel compression is smaller than the QF of cover, the corresponding quantization steps $q''$ will produce a contrary consequence. Also as shown in \reffig{fig-DM}(c):
\begin{equation}\label{equation13}
\tilde d_{q''}\in\lbrack \tilde d-\frac {q''}2,\tilde d+\frac {q''}2\rbrack\;,\;\frac {q''}2>\frac q2\;,\; d\;\neq\;d_{q''},
\end{equation}
where $\tilde d_{q''}$ and $d_{q''}$ are unquantized and quantized coefficients in this situation. Such cases usually involve extraction errors and error propagation due to the utilization of STCs. Here we consider the generalized dither modulation.The above analysis explained the fundamental principle of dither modulation to achieve upward robustness. So dither modulation based algorithms (DMAS and GMAS) require the selection of the image with a lower QF to meet the real-world application scenarios. 

In summary, the embedding using dither modulation is theoretically error-free when the channel compression QF is larger than the cover QF (Upward Robust). But this method still has errors in a real-world implementation even if the QF meets the requirement of Upward Robust. The reason is that we have considered only one of the three lossy processes in the recompression (coefficients quantization). The influences of spatial truncation and rounding were not considered, which will be discussed in the next subsection.


\subsection{Exploration of Spatial Operation}\label{subsection3.2}

According to the inference of \refsec{subsection3.1}, dither modulation-based algorithms are unaffected by \emph{coefficients quantization}. However, when considering the impact of spatial operations, \refequ{equation11} will not necessarily be true. So we will explore this below. 

To measure the effect of the \emph{spatial truncation and rounding}, we summarize the JPEG recompression process in \refsec{notations} as follows,
\begin{equation}\label{equationp1}
\boldsymbol{D}_{q_2}=\lbrack\textbf{DCT}(\lbrack\textbf{TRU}(\textbf{IDCT}(\boldsymbol{D}_{q_1}\times\boldsymbol{Q}_{q_1}))\rbrack)/\boldsymbol{Q}_{q_2}\rbrack.
\end{equation}
First we investigate the impact of spatial rounding. We ignore the truncation operations on the spatial domain. Then the recompression process is represented as follows,
\begin{equation}\label{equationp2}
\begin{aligned}
\boldsymbol{D}_{q_2}&=\lbrack\textbf{DCT}(\lbrack\textbf{IDCT}(\boldsymbol{D}_{q_1}\times\boldsymbol{Q}_{q_1})\rbrack)/\boldsymbol{Q}_{q_2}\rbrack\\
&=\lbrack\textbf{DCT}(\textbf{IDCT}(\boldsymbol{D}_{q_1}\times\boldsymbol{Q}_{q_1})+\boldsymbol{E})/\boldsymbol{Q}_{q_2}\rbrack\\
&=\boldsymbol{D}_{q_2}+\lbrack\textbf{DCT}(\boldsymbol{E})/\boldsymbol{Q}_{q_2}\rbrack\\
&=\boldsymbol{D}_{q_2}+\lbrack\boldsymbol{W}/\boldsymbol{Q}_{q_2}\rbrack,
\end{aligned}
\end{equation}
where $\boldsymbol{E}$ is a rounding error block and all the elements $(e_{i,j})_{8\times8}$ of $\boldsymbol{E}$ belong to $\lbrack-0.5,0.5)$. Here, we assume that $e_{i,j}$ in $\boldsymbol{E}$ is an independent identically distributed (i.i.d.) random variable with uniform distribution in the range of $\lbrack-0.5,0.5)$. Then, based on the central limit theorem, for each element $w_{i,j}$ in $\boldsymbol{W}$, $w_{i,j}$ can be approximately regarded as a Gaussian distribution $\mathcal{N}(0,1/12)$~\cite{JPEG-I}. Consequently, we see that
\begin{equation}\label{equationp3}
\begin{aligned}
P\left\{\left\lbrack\frac{w_{i,j}}{q_{i,j}}\right\rbrack=0\right\}&=P\left\{-\frac{q_{i,j}}{2}\leqslant w_{i,j}<\frac{q_{i,j}}{2}\right\}\\
&\approx\left\{\begin{array}{lc}0.9160,&\text{if}\;q_{i,j}=1\\0.9995,&\text{if}\;q_{i,j}=2\\1,&\text{if}\;q_{i,j}\geqslant3\end{array}\right.,
\end{aligned}
\end{equation}
where $q_{i,j}$ is the quantization step corresponding to the recompression quality factor $q_2$. Therefore, spatial rounding has almost no effect on the recompressed DCT coefficients with recompression $\text{QF}\leqslant92$, since each element of the quantization matrix is larger than 1. This way we can exclude the effect of the \emph{spatial rounding} operation on the robust steganography algorithm. As a conclusion, \textbf{in the case where the recompression QF is larger than the cover QF, \emph{spatial truncation} operation is the primary cause of the error}. Here is an experiment to verify the above conclusion. We characterize the magnitude to which the DCT coefficient is affected by spatial operations as \emph{the stability of the coefficient}. From \reffig{fig-TQ}, we can observe experimentally that truncation operations in the spatial values are the major contributor to the instability of the DCT coefficients, and that rounding operations in the spatial values contribute only relatively small effects.

\begin{figure}[t]
\centering
\includegraphics[width=3.5in]{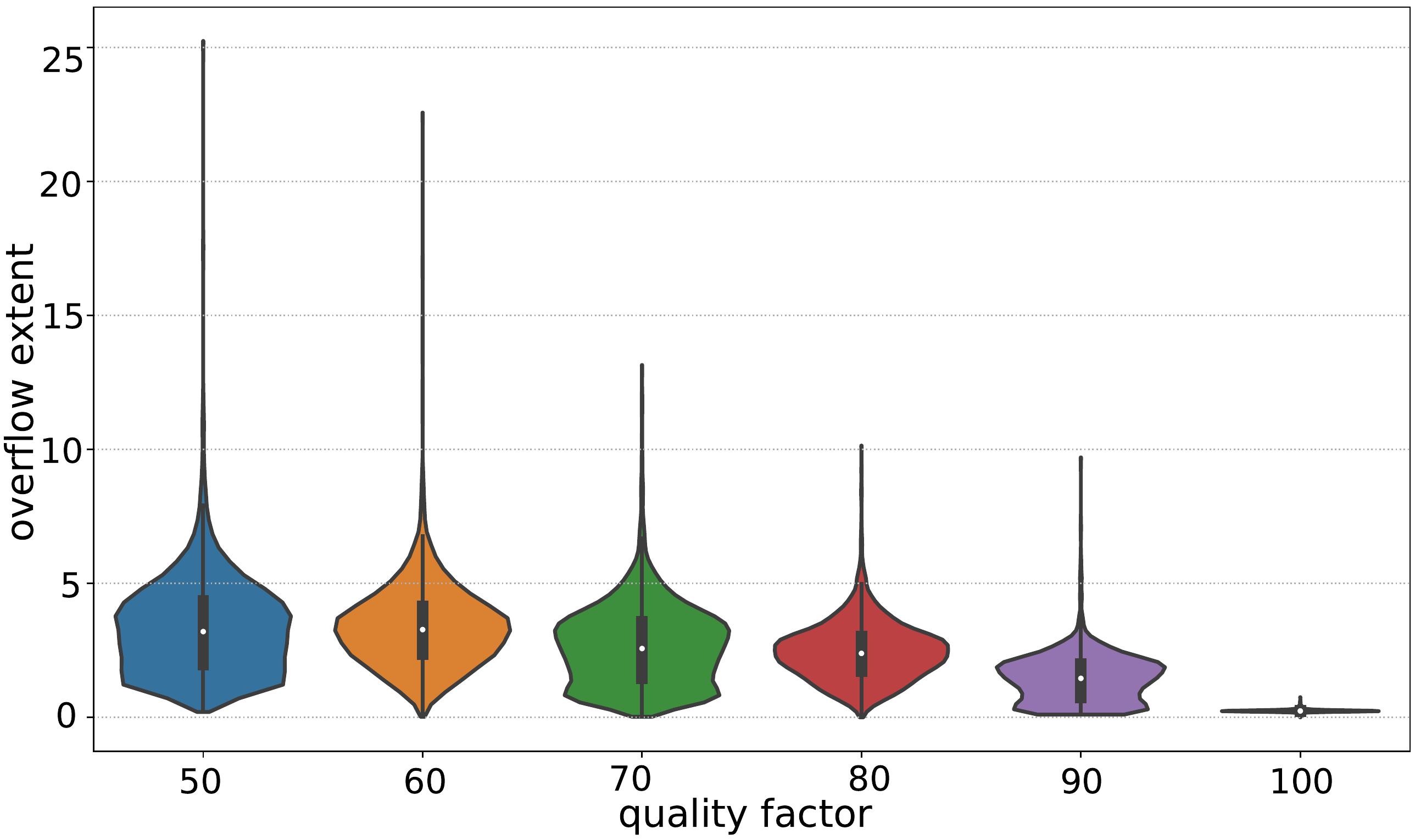}
\caption{\small{The violin chart of overflow extent. We transformed the image in BOSSbase 1.01 \cite{Bossbase} with different QF to the spatial domain and then counted the extent of overflow at different locations. The area of each color map indicates the number of overflow locations.}}
\label{fig-overflow}
\end{figure}

The specific operation of spatial truncation is shown in \refequ{truncation}. Compared to spatial rounding, truncation operations usually cause larger variation, correspondingly also have a greater impact on DCT coefficients, as shown in \reffig{fig-TQ}. Obviously, the exact extent of the impact caused by the truncation operation is related to the magnitude of the corresponding block overflow. And the cause of spatial overflow is that the DCT coefficient derived from the unoverflowed spatial value becomes a coefficient with a larger absolute value during quantization when the image is saved in JPEG format. From the cause of overflow, it is known that the magnitude of overflow depends on the size of the corresponding quantization step. Notably, this quantization step is a save-time one, which is in the quantization table of the cover for the steganographer. We counted the overflow extent of the images in BOSSbase 1.01 \cite{Bossbase} with different QFs and the results are shown in \reffig{fig-overflow}. The smaller the QF of the cover image, the larger the quantization step, the greater the degree of overflow, and the greater the impact suffered from the spatial truncation operation. For robustness, dither modulation based steganography need to choose the image with long quantization step as cover, but this makes them subject to the spatial truncation, even if GMAS chooses to embed in the middle frequency regions with large quantization steps, it is impossible to avoid errors. The reason why repeatedly processing the images through channels can improve the robustness is also that the spatial overflow of the images can be reduced in repeated compression, making the instability of the DCT coefficients diminished. But repeated uploads and downloads are behaviorally insecure. Through the above analysis, we can fundamentally improve the robustness and security by dealing with the image overflow.

\begin{figure}[t]
\centering
\includegraphics[width=3.3in]{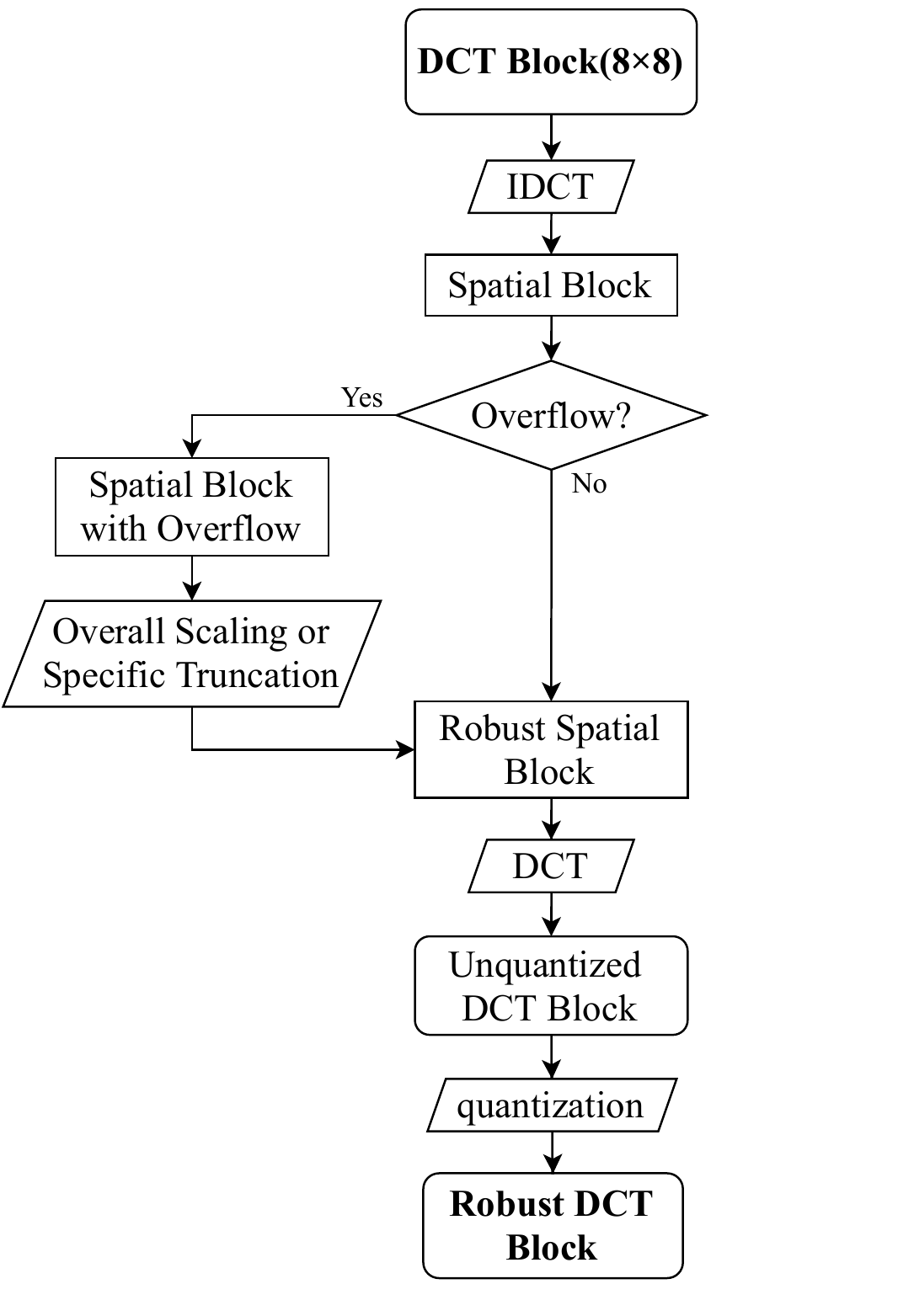}
\caption{\small{The flowchart of removing spatial overflow. The core operations are Overall Scale and Specific Truncation and details of the operation procedures are introduced in \refsec{subsection4.1} and \refsec{subsection4.3}, respectively.}}
\label{fig-recompression}
\end{figure}

In conclusion, \emph{coefficient quantization} results in the primary effect among the three lossy operations, but the application of dither modulation solves this problem. In the two remaining problems, \emph{spatial truncation} plays a major role, while the effect of \emph{spatial rounding} is almost negligible. Consequently, in the procedure of channel compressing, the DCT coefficients of the blocks with the overflow spatial values tend to change dramatically. Inspired by the above discoveries, we have designed a strategy called \textbf{Removing Overflow based Adaptive robuSt sTeganography (ROAST)} to effectively remove the spatial overflow of cover images, which strengthens the stability of the DCT coefficients, thus we do not need to select the robust domain as previous steganography algorithms and can embed on the entire image, which not only significantly enlarges the embedding capacity, but also improves the anti-detection performance remarkably. The proposed strategy ROAST can resist compression primarily because of the removal of spatial overflow, which can be accomplished in two methods: Overall Scale and Specific Truncation. We designed corresponding steganography schemes based on the respective properties of the two approaches. Each is described in detail later.

\section{Proposed Methods}\label{section4}

\begin{figure*}[ht!]
\centering
\includegraphics[width=7.1in]{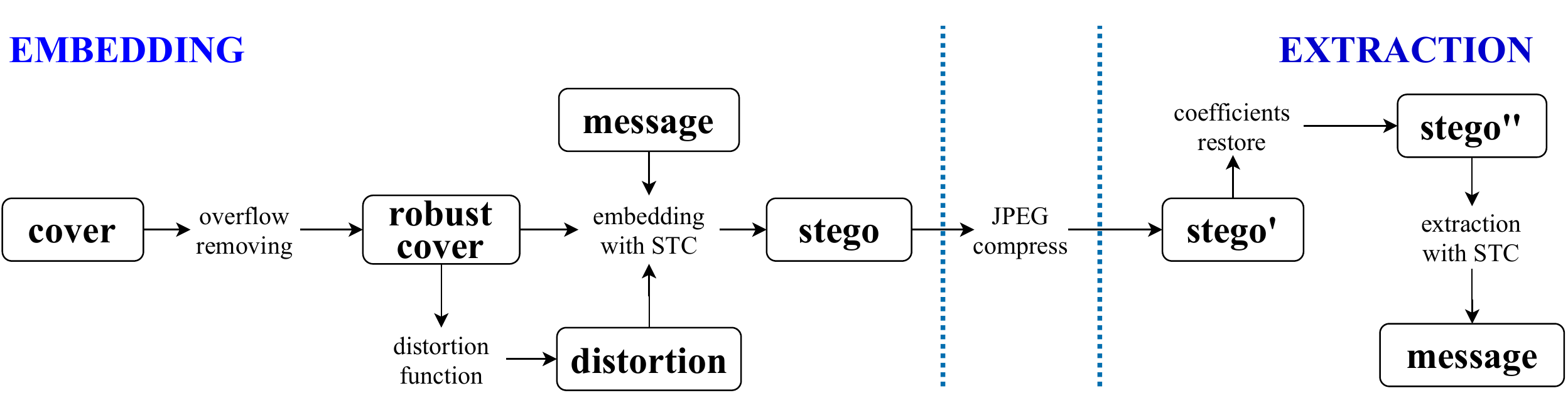}
\caption{\small{Flowchart of the Overall Scale based Steganography (ROAST-OS). The cover, robust cover, and stego are JPEG images and the method of removing overflow is the Overall Scale. Detailed procedures of the algorithm are described in \refsec{subsection4.2}.}}
\label{fig-flowchart2}
\end{figure*}


\subsection{Overall Scale to Remove Overflow}\label{subsection4.1}

The first method of removing spatial overflow is \textbf{Overall Scale (OS)}. This method is primarily concerned with theoretical rigor and enabling greater removal of spatial overflow in particular. The concrete process can be outlined in \reffig{fig-recompression}, and the subsequent part will describe the details and prove the feasibility of this approach.
\begin{enumerate}[i.]
\item Inverse DCT and Inspection. The object of the algorithm is an $8\times8$ DCT coefficient block $\boldsymbol{D}$ of a JPEG image. Convert it to spatial block $\boldsymbol{S}$ with inverse DCT. Denote the unquantized DCT coefficients of row $u$ column $v$ in the DCT block by $\tilde d_{u,v}$ and the spatial values of row $i$ column $j$ in the spatial block by $s_{i,j}$. Note that the spatial value here is not a pixel, an shifting is required to convert it to a pixel. The relationship between the DCT coefficient and the spatial value can be obtained by the formula of DCT:
\begin{equation}\small\label{equation14}
\begin{aligned}
\tilde d_{u,v}\!=\!\frac14\cdot &\boldsymbol{C}(u)\!\cdot\! \boldsymbol{C}(v)\lbrack\sum_{i=0}^7\!\sum_{j=0}^7\!s_{i,j}\!\cdot\!\cos\!\frac{(2i\!+\!1)u\pi}{16}\!\cdot\!\cos\!\frac{(2j\!+\!1)v\pi}{16}\rbrack\;,\\
&\boldsymbol{C}(u)=\left\{\begin{array}{lc}\sqrt{\frac1N}&u=0\\\sqrt{\frac2N}&u\neq0\end{array}\right.,
\end{aligned}
\end{equation}
where $N=8$ denotes the side length of the block. For a more intuitive description, we rewrite \refequ{equation14} to the following form:
\begin{equation}\label{equation15}
\tilde d_{u,v}=\sum_{i,j}s_{i,j}\cdot\Phi_{i,j;u,v}\;,\;i,j,u,v\in\{0,1,\cdots,7\},
\end{equation}
where $\Phi_{i,j;u,v}$ denotes the value of $\tilde d_{u,v}$ when $s_{i,j}=1$, which is the variation caused by a single spatial value on this DCT coefficient. Once the conversion is complete, the overflow can be observed via the spatial value, namely, check if $s_{i,j}$ satisfies $s_{i,j}>127$ or $s_{i,j}<-128$. If there exists an overflow in a block, then this block will be processed to remove the overflow. If not, then the block will be skipped.
\item Scale of Spatial Values. Remove spatial overflow by scaling the whole block to a block without overflow:
\begin{equation}\label{equation16}
s'_{i,j}=\alpha\cdot s_{i,j}\;,\;i,j\in\{0,1,\cdots,7\},
\end{equation}
where $s'_{i,j}$ denotes the spatial value after scaling and $\alpha$ is the parameter. The amount of $\alpha$ depends on the degree of spatial overflow of the block:
\begin{equation}\label{equation17}
\alpha=\frac{s_{\text{0}}}{\left|s_{\left|\text{max}\right|}\right|}\;,\;\;s_{\text{0}}=\left\{\begin{array}{lc}127,&s_{\left|\text{max}\right|}>0\\128\;,&\;else\end{array}\right.,
\end{equation}
where $s_{\left|\text{max}\right|}$ denotes the spatial value with the maximum absolute value. So that $\alpha$ calculated in this way makes the scaled block without overflow.
\item DCT and Quantization. Convert the block which has been removed overflow into DCT coefficients. The DCT coefficients corresponding to the scaled block are $\tilde d'_{u,v}$:
\begin{equation}\label{equation18}
\tilde d'_{u,v}=\sum_{i,j}s'_{i,j}\cdot\Phi_{i,j;u,v}\;=\alpha\sum_{i,j}s_{i,j}\cdot\Phi_{i,j;u,v}\;=\alpha\cdot \tilde d_{u,v}.
\end{equation}
At this time, this block gets rid of overflow. Thereafter, the DCT coefficients will be quantized to integer.
Generally, the round function “$\left[\;\cdot\;\right]$” is adopted which modifies a number to the nearest integer for quantization. However, the round operation will cause undesirable changes, e.g. the coefficient increases. The former scaling changes intend to reduce the amplitude of DCT coefficients, therefore \textbf{FIX}$(\;\cdot\;)$ which modifies the number to the nearest integer with a smaller absolute value is utilized here for quantization.
\begin{equation}\label{fix}
\textbf{FIX}(x)=\left\{\begin{array}{lc}\left\lfloor x\right\rfloor&x\geqslant 0\\\left\lceil x\right\rceil& else\end{array}\right..
\end{equation}
The final processed DCT coefficient is $d_{u,v}^*$,
\begin{equation}\label{processed}
d_{u,v}^*=\textbf{FIX}(\tilde d'_{u,v}/q_{u,v}),
\end{equation}
where $q_{u,v}$ is the corresponding quantization step. 
\end{enumerate}

Here gives the proof that the processed block is without the overflow. The unquantized DCT coefficient corresponding to the finally stored DCT coefficient is $\tilde d_{u,v}^*$:
\begin{equation}\label{equation19}
\tilde d_{u,v}^*=d_{u,v}^* \cdot q_{u,v} = \textbf{FIX}(\tilde d'_{u,v}/q_{u,v}) \cdot q_{u,v}=\beta_{u,v}\cdot \tilde d_{u,v},
\end{equation}
where $\beta_{u,v}$ is a parameter jointly determined by quantization and overall scale. It is obvious from the above process that $\beta_{u,v}\leqslant\alpha$. Similar to the way the DCT formula in \refequ{equation15}, rewrite the inverse DCT in the following format:
\begin{equation}\label{equation20}
s_{i,j}=\sum_{u,v} \tilde d_{u,v}\cdot\Psi_{u,v;i,j},
\end{equation}
where $\Psi_{u,v;i,j}$ denotes the value of $s_{i,j}$ when $d_{u,v}=1$. The spatial value corresponding to the adjusted coefficients is $s_{i,j}^*$ :
\begin{equation}\label{equation21}
s_{i,j}^*=\sum_{u,v} \tilde d_{u,v}^*\cdot\Psi_{u,v;i,j}=\;\sum_{u,v}\beta_{u,v}\cdot \tilde d_{u,v}\cdot\Psi_{u,v;i,j}.
\end{equation}
Apparently, $s_{i,j}^*\leqslant s'_{i,j}$ as a result of $\beta_{u,v}\leqslant\alpha<1$. Hence, the corresponding spatial values diminish spatial overflow after adjusting the DCT coefficients. Even though it may occasionally cause overflow after steganography, the effect will be minimal compared to before.

\subsection{Overall Scale based Steganography: ROAST-OS}\label{subsection4.2}

The stability of the coefficients can be significantly improved via the Overall Scale method introduced in the previous section, so we use this method and generalized dither modulation to design an upward robust steganography scheme named ROAST-OS. \reffig{fig-flowchart2} illustrates the procedure of the scheme. The process and details are described hereunder:
\begin{enumerate}[i.]
\item Preprocessing of Cover Image. Removing the spatial overflow of cover image by the Overall Scale method. The cover after preprocessing is generally considered to be sufficiently robust that the whole image can be regarded as a robust region, which we called \textbf{robust cover}.
\item Define Distortion for Robust Cover. Define the asymmetric distortion of robust cover using the \refequ{equation6} and \refequ{equation7}. Here the distortion of the quantized DCT coefficients instead of the unquantized DCT coefficients are calculated, that is, \refequ{equation8} and \refequ{equation9} are not required here. This is according to the \refequ{equation10} which means the unquantized DCT coefficients of the JPEG image are exactly at the center of the region divided by the quantization step.
\item RS Encoding and Ternary STC Embedding. Utilizing RS codes to encode the messages and ternary STC to embed.
\end{enumerate}

The receiver first performs the DCT coefficients restoration introduced in \refsec{subsection2.2} on the compressed stego image. The message is then decoded with the RS code and STCs.

\subsection{Specific Truncation to Remove Overflow}\label{subsection4.3}

\begin{figure*}[t]
\centering
\includegraphics[width=7.1in]{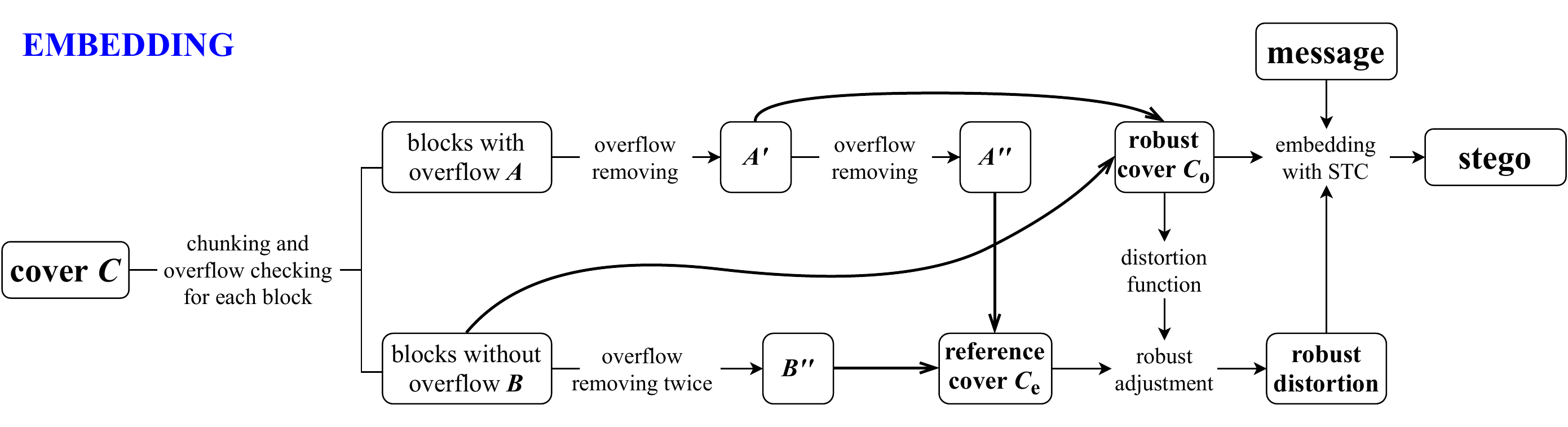}
\caption{\small{Flowchart of the Specific Truncation based Steganography (ROAST-ST). The scheme consists of three components: preprocessing, calculate robust asymmetric distortion, and STCs Embedding. Detailed procedures of the algorithm are described in \refsec{subsection4.4}. Note that the overflow blocks here refer to the blocks that satisfy the condition proposed in \refsec{subsection4.4}.}}
\label{fig-flowchart}
\end{figure*}

Although the aforesaid Overall Scale method can remove the overflow to produce a robust cover and is also outstanding in practical steganography schemes, we need to consider the anti-detection performance while pursuing robustness in robust steganography. The employment of the Overall Scale generates a large number of modifications, which greatly affects the anti-detection performance, so we devised a method named \textbf{Specific Truncation (ST)} for targeted tuning of the overflow location and parameterizing the processing intensity to remove spatial overflow with relatively fewer modifications. The concrete procedure of the method is shown in \reffig{fig-recompression}, and the details are presented as follows.
\begin{enumerate}[i.]
\item Inverse DCT and Inspection. Convert the DCT block to spatial values and detect whether the block is overflowing or not, namely, check if there is $s_{i,j}$  satisfying $s_{i,j}>127$ or $s_{i,j}<-128$. If there exists an overflow, then the block will be processed to remove the overflow. If not, then it will be ignored.
\item Specific Truncation of Overflow Values. For all out-of-range positions, denoted as $\hat{s}_{i,j}$ ($\hat{s}_{i,j}>127$ or $\hat{s}_{i,j}<-128$), the truncation operation is performed as follows:
\begin{equation}\label{equation22}
\hat{s}_{i,j}=\left\{\begin{array}{lc}127-T_1,&\hat{s}_{i,j}>127\\-128+T_1,&\hat{s}_{i,j}<-128\end{array}\right.,
\end{equation}
$T_{1}$ is a parameter to control the intensity of the truncation operation and the magnitude of the modification, which pursues a better trade-off between robustness and resistance to detection. Denote the processed spatial block as $\boldsymbol{\widehat S}$.
\item DCT and Quantization. Convert the processed spatial block $\boldsymbol{\widehat S}$ to coefficients $\boldsymbol{\widetilde D}$ using DCT:
\begin{equation}\label{equation23}
\boldsymbol{\widetilde D}=\textbf{DCT}(\boldsymbol{\widehat S}).
\end{equation}
Following the setting in section \refsec{subsection4.1}, we apply \textbf{FIX}() in the quantization phase as well.
\begin{equation}\label{equation24}
\boldsymbol{D}=\textbf{FIX}(\boldsymbol{\widetilde D}/\boldsymbol{Q})=\textbf{FIX}(\textbf{DCT}(\boldsymbol{\widehat S}) / \boldsymbol{Q}),
\end{equation}
$\boldsymbol{D}$ is the robust block after the removal of the spatial overflow.
\end{enumerate}

\subsection{Specific Truncation based Steganography: ROAST-ST}\label{subsection4.4}

Specific Truncation is a de-overflow method with relatively smaller modifications compared to Overall Scale, but it also unavoidably weakens the performance of overflow removing. The steganographic scheme called ROAST-ST is designed to enhance the robustness by using robust asymmetric distortion. As shown in \reffig{fig-flowchart}, it consists of three parts: preprocessing, calculate robust asymmetric distortion, and STCs embedding. Each step of the scheme is described in detail below:
\begin{enumerate}[i.]
\item Preprocessing. The preprocess is designed to enhance the stability of the DCT coefficients by removing the spatial overflow with Specific Truncation operations. When determining whether a block is overflowed or not, it is not as straightforward as before, but instead first calculates the extent of the overflow. Because the preprocessing always has a more significant impact on the image compared to the embedding process and an intuitive way to improve security is to make fewer changes. The overflow of a block is measured by $\Omega$, which is calculated as follow:
\begin{equation}\label{equation25}
\Omega=\sum_{i,j}\delta_{i,j}\;,\;\delta_{i,j}=\left\{\begin{array}{lc}\left|s_{i,j}\right|-127\;,&s_{i,j}>127\;\\\left|s_{i,j}\right|-128\;,&s_{i,j}<-128\\0\;,&else\end{array}\right..
\end{equation}
Actually, it is the de-overflow operation rather than steganographic modification impacts on the security of the block. We therefore introduce $T_2$ to control whether or not the block is considered overflow. The block will be considered as overflow block if $\Omega>T_2$ or not if $\Omega \leq T_2$. This way we can divide the blocks of cover $\boldsymbol{C}$ into two groups: $\boldsymbol{A}$ with overflow and $\boldsymbol{B}$ without overflow.
A de-overflow operation on $\boldsymbol{A}$ yields $\boldsymbol{A'}$ and $\boldsymbol{B}$ doesn't do the processing, thus obtain a robust cover $\boldsymbol{C_{\text{o}}}$ for steganography. While reducing the magnitude of the modification in the Specific Truncation, it also leads to the consequence that it is difficult to completely remove all spatial overflow with this method used only once. So we expect to remove spatial overflow further during embedding. A de-overflow operation on $\boldsymbol{A'}$ again yields $\boldsymbol{A''}$ and twice de-overflow operations on $\boldsymbol{B}$ yields $\boldsymbol{B''}$, thus obtain a reference cover $\boldsymbol{C_{\text{e}}}$ with less overflow than robust cover. Making $\boldsymbol{C_{\text{o}}}$ close to $\boldsymbol{C_{\text{e}}}$ in the process of embedding modifications can enhance coefficient stability. 
\item Define Robust Asymmetric Distortion. In the aforementioned preprocessing, not only a robust cover is prepared, but also a reference image is given for designing asymmetric distortion. Therefore we can combine the de-overflow into the process of message embedding. In detail, we employ the distortions $\rho_{i,j}^+$ and $\rho_{i,j}^-$ calculated by the \refequ{equation6} and \refequ{equation7} on the robust cover as the basic distortions. Denote the quantized DCT coefficient of the robust cover and the reference image as $d_{i,j}^{\text{o}}$ and $d_{i,j}^{\text{e}}$, respectively. Robust asymmetric distortion is calculated as follows:
\begin{equation}\label{equation26}
\varepsilon_{i,j}^+=\left\{\begin{array}{lc}\mu\cdot\rho_{i,j}^+,&d_{i,j}^{\text{e}}>d_{i,j}^{\text{o}}\\\rho_{i,j}^+,&else\end{array}\right.,
\end{equation}
\begin{equation}\label{equation27}
\varepsilon_{i,j}^-=\left\{\begin{array}{lc}\mu\cdot\rho_{i,j}^-,&d_{i,j}^{\text{e}}<d_{i,j}^{\text{o}}\\\rho_{i,j}^-,&else\end{array}\right.,
\end{equation}
$\varepsilon_{i,j}^+$ and $\varepsilon_{i,j}^-$ are asymmetric distortion after performing robustness adjustment. $0<\mu\leqslant1$ is a parameter for adjusting distortion. This defines a smaller ``+1" cost for locations where the coefficient of the reference image is larger than that of the robust cover and a smaller ``-1" cost for locations where the coefficient of the reference image is less than that of the robust cover.
\item RS Encoding and Ternary STC Embedding. This part is the same as mentioned before.
\end{enumerate}

The receiver extracts the message in the same way as \refsec{subsection4.2}, which is not repeated here.

\section{Experiment}\label{section5}

\subsection{Setups}\label{subsection5.1}

All experiments in this paper are conducted on BOSSbase 1.01~\cite{Bossbase} containing 10,000 grayscale $512 \times 512$ images. We randomly selected 2000 of them and the original images are JPEG compressed using quality factor 65 as the cover image. From \refsec{section3} we know that the dither modulation based algorithm needs to choose a cover with a relatively low QF to achieve robustness. Therefore, we use 65 as an example in our experiments. The relative payload is $n_\textrm{m} / n_\textrm{nzac}$, where $n_\textrm{m}$ is the length of the original embedded messages rather than the encoded messages by RS codes and $n_\textrm{nzac}$ is the number of non-zero AC DCT coefficients of the original cover image rather than the robust cover after de-overflow processing. Previous steganographic algorithms typically set the range of the relative payloads of robust adaptive steganography from $0.05$ to $0.15$ bits per non-zero AC DCT coefficients (bpnzac) due to their limited performance. We refer to such payloads as low payloads and perform experiments with high payload from $0.1$ to $0.5$ (bpnzac) in this paper. The extraction error rate $R_\textrm{error}=n_\textrm{error} / n_\textrm{m}$, where $n_\textrm{error}$ is the number of wrong message bits. The QF of the cover image and channel JPEG compression is denoted as $Q_\textrm{cover}$ and $Q_\textrm{channel}$, respectively. Two effective feature sets (CCPEV~\cite{CCPEV}, DCTR~\cite{DCTR}) and SRNet~\cite{SRNet} are selected for steganalysis of JPEG image. We will set the secure parameter $h=10$ of STCs, and RS (31,15) will be adopted in the following experiments. As for ($n^*$, $k^*$) RS codes, $n^*$ and $k^*$ denote the code length and message length respectively, and the greater the ratio of $k^*$ to $n^*$, the stronger the error correction ability of RS codes.

The detectors are trained as binary classifiers implemented using the FLD ensemble with default settings~\cite{ensemble}. The ensemble by default minimizes the total classification error probability under equal priors $P_\textrm{E} = \min_{P_{\textrm{FA}}}\frac{1}{2}(P_{\textrm{FA}}+P_{\textrm{MD}})$, where $P_{\textrm{FA}}$ and $P_{\textrm{MD}}$ are the false-alarm probability and the missed-detection probability respectively. SRNet is fed the cover and stego images. The training first runs for 300 epocks with an initial learning rate of $r_1=0.001$ and then for an additional 100 epochs with a learning rate of $r_2=0.0001$. A separate classifier and a deep network are trained for each embedding algorithm and payload. The ultimate security is qualified by average error rate $\overline{P}_\textrm{E}$, and larger $\overline{P}_\textrm{E}$ means stronger security.

\subsection{Investigation of the Preprocessing}\label{subsection5.2}

The preprocessing improves the stability of the DCT coefficients by removing the spatial overflow of the cover image. It is an operation that has a relatively high impact on security and robustness therefore we use parameters to find a trade-off between them. Subsequently, we will explore these in detail through experiments.

\subsubsection{Security of Preprocessing}\label{subsection5.2.1}

\begin{figure}[t]
\centering
\includegraphics[width=3in]{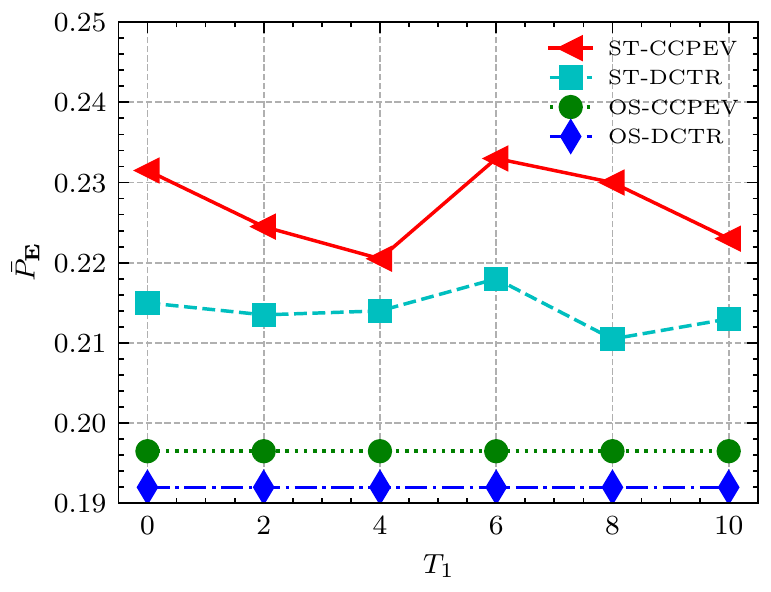}
\caption{\small{Comparison of two methods Overall Scale (OS) and Specific Truncation (ST) for removing spatial overflow on anti-detection performance as well as the effect of parameter $T_1$ on Specific Truncation ($Q_\textrm{cover}=65$).}}
\label{fig-T1_CCPEV_DCTR_s}
\end{figure}

Both methods of removing overflow in the preprocessing modify the original images. Although the final steganography process is performed on the modified images, the attacker is likely to use the original image for training in steganalysis. It generally seems to us that the original image should be more readily available. In contrast to~\cite{TCM}, this paper proposes that the \textbf{practical security} of robust steganography should be evaluated by the original image before processing and the image after steganography.

The effect of the preprocessing process on steganography security is shown in \reffig{fig-T1_CCPEV_DCTR_s}. The error detection rate in the figure is detected on the preprocessed image when the payload is 0. The horizontal coordinates are about the parameter $T_1$ in Specific Truncation which has no influence on the Overall Scale. From the experimental results, it can be seen that the preprocessing operation has a sharp decrease in anti-detection performance. The Specific Truncation has less decrease on the anti-detection ability than the Overall Scale method. Additionally, the change in the parameter $T_1$ has little influence on the anti-detection performance of ROAST-ST. The above consequences are consistent with expectations because of a large number of coefficients modified during preprocessing and there are relatively slight number of points modified as $T_1$ changes.

\subsubsection{Effect of Preprocessing on Robustness}\label{subsection5.2.2}

\begin{figure}[t]
\centering
\includegraphics[width=3in]{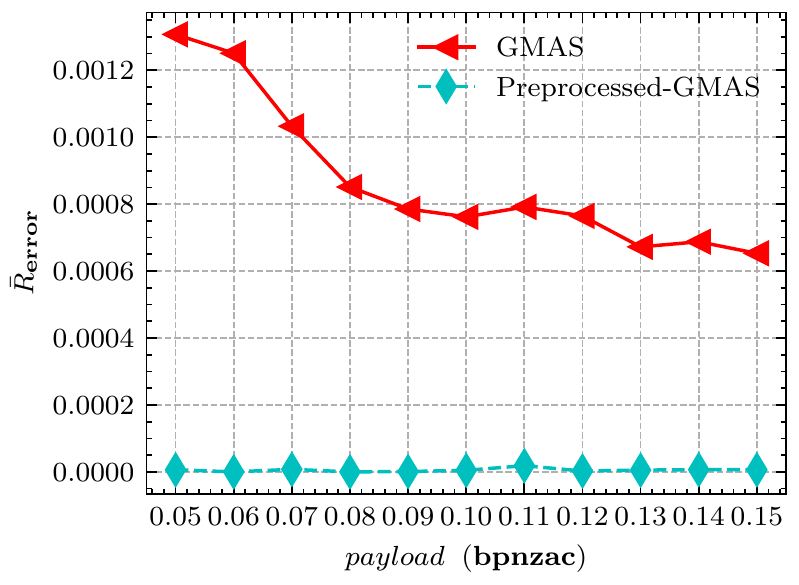}
\caption{\small{Comparison of compression resistance between applying GMAS on images preprocessed with the Specific Truncation method ($T_1=8$) and applying GMAS on the original images ($Q_\textrm{cover}=65$ and $Q_\textrm{channel}=85$). $\overline{R}_\textrm{error}$ is average extraction error rate.}}
\label{fig-preprocessed-GMAS_GMAS_r}
\end{figure}

To demonstrate the improvement in the anti-compression capability of the preprocessing operation, we have compared the robustness of employing GMAS on images preprocessed by the Specific Truncation method ($T_1=8$) with that of employing GMAS directly on the original images. The experimental results presented in \reffig{fig-preprocessed-GMAS_GMAS_r} show that the preprocessing operation has a significant effect on improving the stability of the coefficient and the robustness of the algorithm. It can also be noted that combining the preprocessing with the GMAS algorithm can virtually make $R_{error}=0$ at low payloads, indicating that the preprocessing does make sense for improving robustness. It is clear from \reffig{fig-T1_CCPEV_DCTR_s} and \reffig{fig-preprocessed-GMAS_GMAS_r} that preprocessing can dramatically improve robustness at the sacrifice to security through extensive modifications. So we use $T_1$ and $T_2$ to make a trade-off. The effect of each parameter on the robustness is shown below.

\subsubsection{Effect of Parameter $T_1$ on Robustness}\label{subsection5.2.3}

\begin{figure}[t]
\centering
\includegraphics[width=3in]{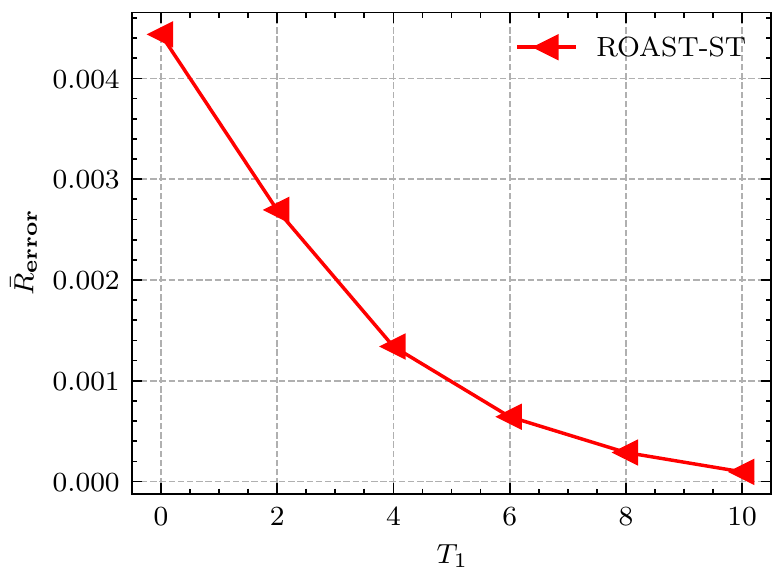}
\caption{\small{The influence of the parameter $T_1$ in the Specific Truncation method on the robustness of ROAST-ST ($Q_\textrm{cover}=65$ , $Q_\textrm{channel}=85$ , $payload=0.3$).}}
\label{fig-UPRS-ST_T1_R}
\end{figure}

In \refsec{subsection5.2.1}, we discovered that changes of $T_1$ in the preprocessing have an ignorable effect on the anti-detection performance because the additional modifications due to parameter variations are marginal. However, changes in $T_1$ can have a tangible effect on the robustness of the ROAST-ST as shown in \reffig{fig-UPRS-ST_T1_R}. As $T_1$ increases, the robustness of the ROAST-ST algorithm improves considerably, and when $T_1\geqslant8$, it becomes more robust than the ROAST-OS algorithm. Utilization of $T_1$ can enhance robustness with little impact on security. This phenomenon suggests that the preprocessing of the algorithm is effective in conjunction with the robust asymmetric distortion design.

\subsubsection{Effect of Parameter $T_2$}\label{subsection5.2.4}

\begin{figure*}[t]
\centering\subfigure[Security]{
\includegraphics[width=3in]{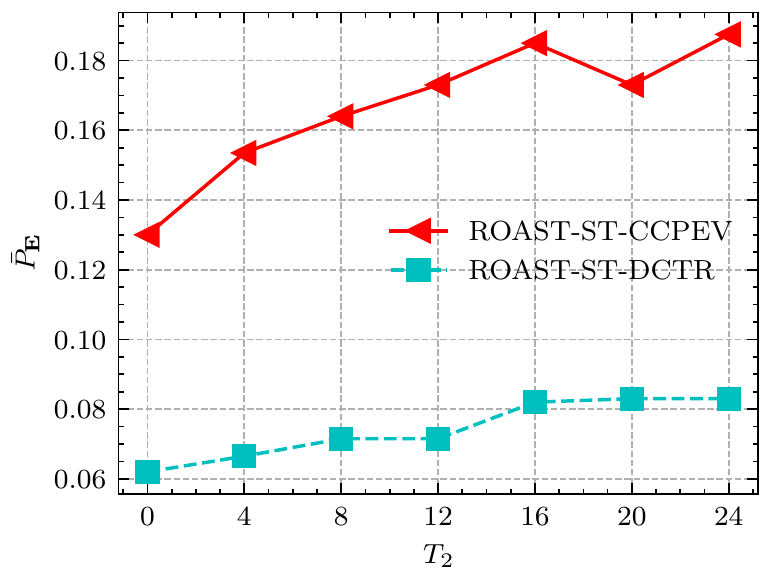}}
\centering\subfigure[Robustness]{
\includegraphics[width=3in]{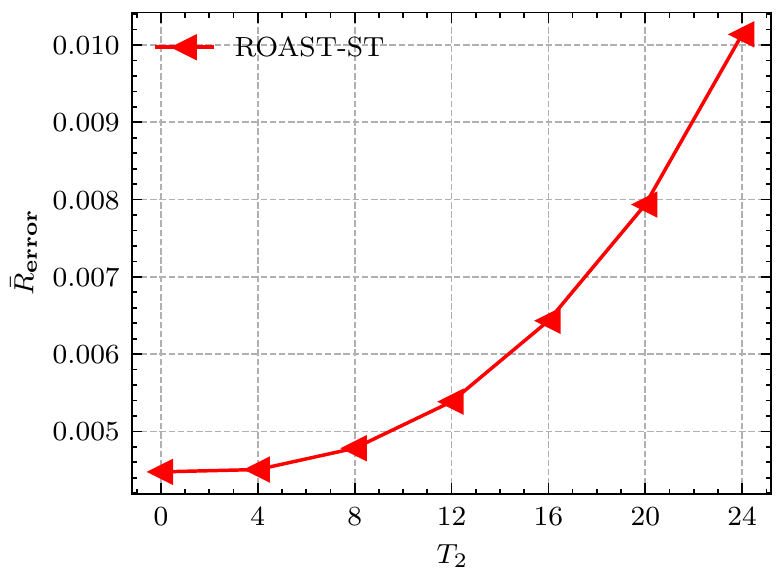}}
\caption{\small{Effect of Parameter $T_2$ in ROAST-ST on (1) security and (2) robustness ($Q_\textrm{cover}=65$ , $Q_\textrm{channel}=85$ , $payload=0.3$ , $T_1=0$).}}
\label{fig-T2}
\end{figure*}

The parameter $T_2$ is the threshold for determining whether preprocessing is performed. See \refsec{subsection4.4} for details and \reffig{fig-T2} for experimental results. Obviously, as $T_2$ increases, the number of modified blocks decreases, the robustness of the ROAST-ST algorithm decreases and the security increases. This confirms the previous statement and makes the algorithm more flexible to adjust security and robustness to achieve practical requirements.

\subsection{Investigation of Robust Asymmetric Distortion}\label{subsection5.3}

\subsubsection{Effect of Robust Asymmetric Distortion on Robustness}\label{subsection5.3.1}

\begin{figure}[t]
\centering
\includegraphics[width=3in]{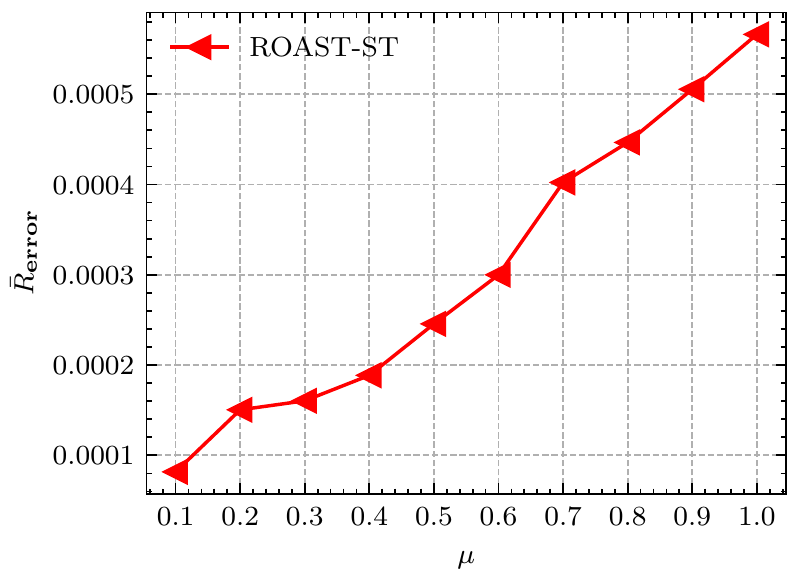}
\caption{\small{The influence of the parameter $\mu$ in the ROAST-ST on the robustness ($Q_\textrm{cover}=65$ , $Q_\textrm{channel}=85$ , $payload=0.3$ , $T_1=8$ , $T_2=0$).}}
\label{fig-UPRS-ST_mu_R}
\end{figure}

In this section, we examine the effect of robust asymmetric distortion on robustness. This asymmetric distortion concept is analogous to the side information steganography~\cite{SI}, in which the side information is used to improve security, while this asymmetric distortion is used to improve robustness. Therefore, we adopt the image after the second de-overflow operation as the reference image. The exact calculation of distortion is described in detail by \refsec{subsection4.4}. The robustness of the ROAST-ST, which employs asymmetric distortion, varies with $\mu$ as shown in \reffig{fig-UPRS-ST_mu_R}. As the effect of asymmetric distortion increases with decreasing $\mu$, the robustness of the ROAST-ST is considerably improved. The robustness of the ROAST-ST is comparable to that of the ROAST-OS at $\mu=0.5$, and stronger robustness is obtained at a smaller $\mu$. This means that the definition of robust asymmetric distortion indeed contributes to the robustness. Notably, the robust distortion adjustment only adjusts a relatively small amount of the distortion, so even if $\mu$ is chosen to be very small, the impact on security is also acceptable. This will be discussed in the next section.

\subsubsection{Security of Robust Asymmetric Distortion}\label{subsection5.3.2}

\begin{figure}[t]
\centering
\includegraphics[width=3in]{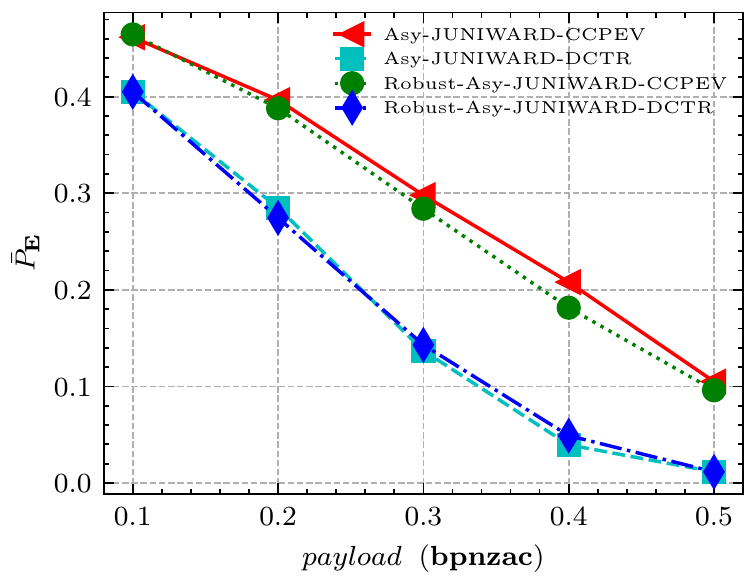}
\caption{\small{Security comparison between basic asymmetric distortion and robust asymmetric distortion. Only images before and after steganography are used here for steganalysis experiments to avoid interference of preprocessing ($Q_\textrm{cover}=65$ , $Q_\textrm{channel}=85$ , $\mu=0.5$ , $T_1=8$ , $T_2=0$).}}
\label{fig-UPRS-ST_mu_s}
\end{figure}

The above experiments demonstrate that the proposed asymmetric distortion can effectively improve robustness. To investigate the role of robust adjustment on security, we compare the security between the asymmetric distortion in GMAS and the robust asymmetric distortion proposed in this paper. The covers utilized here are all images processed by the Specific Truncation method ($T_1=8$). The experimental results and concrete parameters are shown in \reffig{fig-UPRS-ST_mu_s}. Only images before and after steganography are used here for steganalysis experiments to avoid interference of preprocessing. As can be seen, robustness adjustments do cause a decrease in anti-detection performance. However, the  difference of security is a trivially visible difference when using feature CCPEV for steganalysis, and even more trivial when using feature DCTR. An essential observation from previous experiments is that security and robustness cannot always be achieved perfectly at the same time. There will always be some sacrifices to security involved in improving robustness. Overall, the impact on security is still acceptable when applying parameters and asymmetric distortion to enhance robustness. This is the reason ROAST-ST has an advantage over ROAST-OS in terms of both security and robustness at lower payloads. The following is a detailed demonstration. 

\subsection{Comparison with the State-of-the-art Algorithms}\label{subsection5.4}

\begin{figure*}[t]
\centering\subfigure[$Q_\textrm{channel}=85$]{
\includegraphics[width=3in]{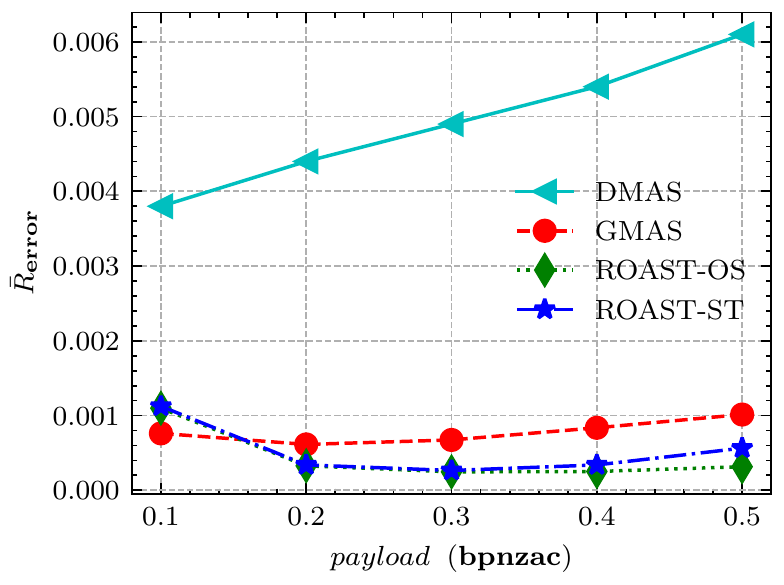}}
\centering\subfigure[$Q_\textrm{channel}=95$]{
\includegraphics[width=3in]{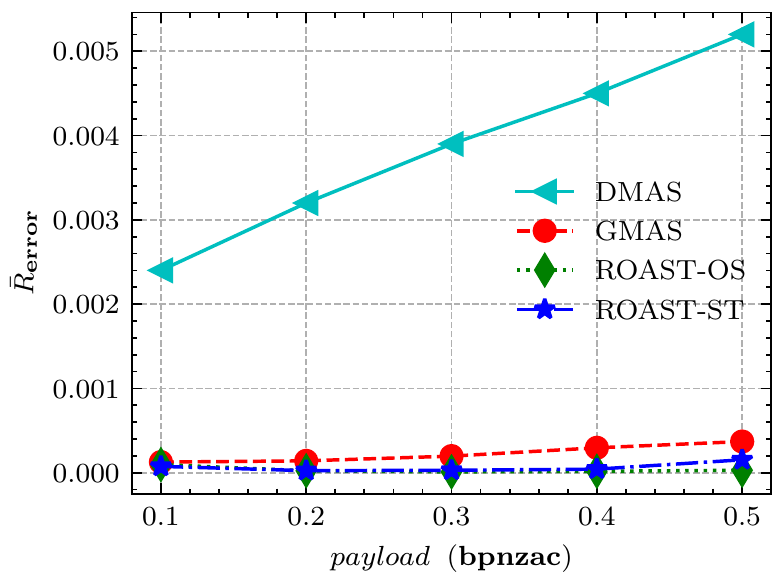}}
\caption{\small{Comparison of the robustness among various algorithms at (1) $Q_\textrm{channel}=85$ and (2) $Q_\textrm{channel}=95$ ($Q_\textrm{cover}=65$ , $T_1=8$ , $T_2=0$ , $\mu=0.5$).}}
\label{fig-h-UPRS-ST_GMAS_UPRS-OS_r}
\end{figure*}

\begin{figure*}[t]
\centering\subfigure[CCPEV]{
\includegraphics[width=3in]{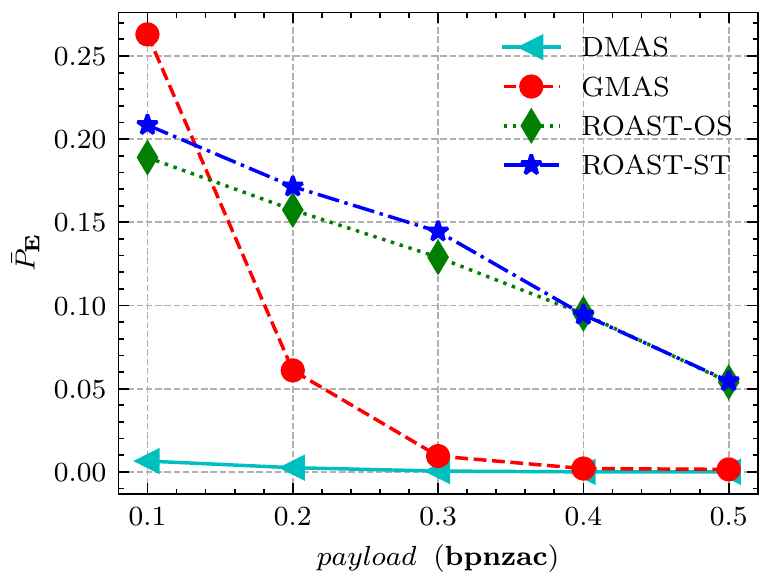}}
\centering\subfigure[DCTR]{
\includegraphics[width=3in]{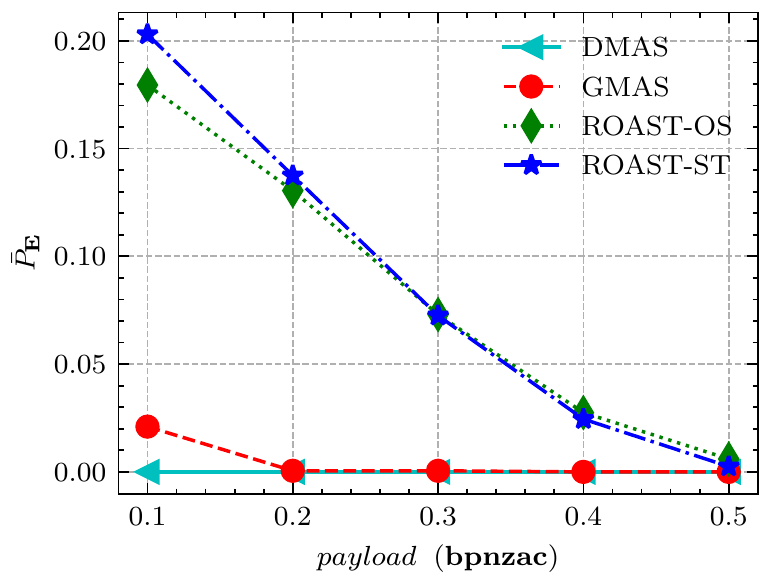}}
\caption{\small{Comparison of the security among various algorithms utilizing (1) CCPEV and (2) DCTR as steganalysis features ($Q_\textrm{cover}=65$ , $T_1=8$ , $T_2=0$ , $\mu=0.5$).}}
\label{fig-85-h-UPRS-ST_GMAS_UPRS-OS_s}
\end{figure*}

\begin{figure}[t]
\centering
\includegraphics[width=3in]{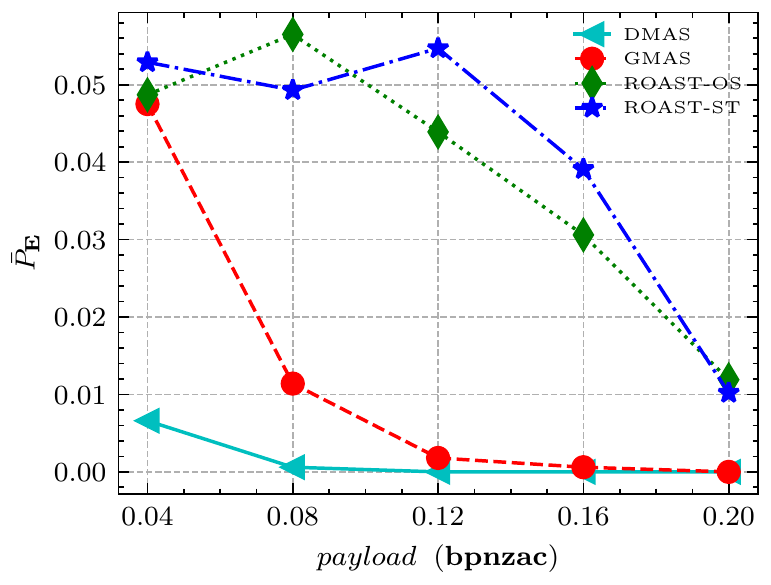}
\caption{\small{Comparison of the security among various algorithms utilizing SRNet ($Q_\textrm{cover}=65$ , $T_1=8$ , $T_2=0$ , $\mu=0.5$).}}
\label{fig-85-h-UPRS-ST_GMAS_UPRS-OS_s_SRNet}
\end{figure}


In this section, we compare the proposed algorithms with state-of-the-art algorithms. First of all, in terms of applicability, the algorithm proposed by Tao \emph{et al.}~\cite{Downward-robust} only works in the scenario that the QF of channel compression is less than that of cover images (Downward Robust). TCM~\cite{TCM} works well only if the channel parameters are known or if the channel can be used at will. Therefore, we compare the proposed algorithms to GMAS~\cite{GMAS} and DMAS~\cite{DMAS} that have the same application environment. After that, we choose parameters for ROAST-ST that make it perform comparable to ROAST-OS according to the results of the previous experiment. The specific values are $T_1=8$ , $T_2=0$ , $\mu=0.5$.


The comparison of robustness is shown in \reffig{fig-h-UPRS-ST_GMAS_UPRS-OS_r} and the security is shown in \reffig{fig-85-h-UPRS-ST_GMAS_UPRS-OS_s}. In terms of robustness, except for $Q_{\text{channel}}=85$ with payloads near $0.1$ bpnzac, both algorithms proposed in this paper have significant advantages over the previous algorithms. Comparing robustness at $Q_{\text{channel}}=85$ and $Q_{\text{channel}}=95$, we can obtain that as the recompression QF decreases, the robustness performance of the algorithm decreases. This is consistent with the inference in \refsec{subsection3.1} regarding the recompression QF. In terms of security, it is also the proposed algorithms that have significant advantages, except when using CCPEV feature and the payload is near $0.1$ bpnzac. Representatively, average detection error rate $\overline P_\textrm{E}$ is increased by +18\% when utilizing DCTR features with $payload=0.1$ bpnzac and +14\% when utilizing CCPEV features with $payload=0.3$ bpnzac. It is worth mentioning that both DMAS and GMAS almost fail to resist detection at $payload\geqslant0.2\;\;(\text{bpnzac})$ when utilizing more powerful DCTR features, and the proposed algorithms are even more outstanding in this situation. We also compare the security when using the popular CNN-based steganalysis in \reffig{fig-85-h-UPRS-ST_GMAS_UPRS-OS_s_SRNet}. The security of the proposed method is enhanced at low embedding rate compared with DMAS and GMAS. Average detection error rate $\overline P_\textrm{E}$ is increased by +5\% when utilizing SRNet with $payload=0.12$ bpnzac. Overall ROAST-ST and ROAST-OS achieve notable robustness and security.

The two algorithms proposed in this paper have their strengths and weaknesses, due to the use of de-overflow operations and corresponding steganographic strategies. From \reffig{fig-85-h-UPRS-ST_GMAS_UPRS-OS_s}, roughly divided by $payload=0.3\;\;(\text{bpnzac})$, ROAST-ST has superiority in both robustness and security when the payload is small, and ROAST-OS has superiority in both when the payload is large. With smaller payloads, the security of the algorithm is substantially determined by the method of preprocessing, so this case can demonstrate the benefits of fewer preprocessing modifications and robust asymmetric distortions in ROAST-ST. When the payload is large, the modifications caused by embedding also progressively take effect, and therefore the strengths of fewer modifications in preprocessing operations are not so much, instead, the stable robustness of more preprocessing modifications and the security of distortion without robust adjustment will be exhibited.  Consequently, both of these two algorithms are characterized by their properties.

\section{Conclusions}\label{section6}

Robust steganography is a research dedicated to the practical application of steganography. JPEG compression is the current primary processing operation for user-uploaded images in social networks, which is also the major focus of robust steganography research.

This paper explores the effect of JPEG recompression on the dither modulation algorithms by theoretical derivation based on DMAS and GMAS. Then we propose the critical steps required to implement effective upward robust steganography: the removal of spatial overflow. The advanced robust steganography algorithms are designed accordingly. The proposed algorithm enhances the stability of coefficients by eliminating overflow thus does not require the selection of a robust domain when embedding message and greatly improves robustness and security. This also enables a much higher embedding capacity than previous algorithms. Besides the concise steganography scheme ROAST-OS, we design a more practical scheme ROAST-ST by utilizing asymmetric distortion. ROAST-ST can improve the detection resistance by reasonably reducing the modification magnitude while ensuring robustness. Based on this, we discuss the dichotomy between robustness and security as well as how to make a trade-off or find a reasonable balance. Even though the algorithm in this paper has been significantly improved in performance compared to the previous algorithm, it is still necessary to reasonably adjust the performance for the practical scenario under the defined performance limits, which is why we have designed multiple parameters. In the future, we will improve the method to more realistic and complex scenarios.
\section*{Acknowledgment}

The authors would like to thank DDE Laboratory of SUNY Binghamton for sharing the source code of steganography, steganalysis and ensemble classifier on the webpage (http://dde.binghamton.edu/download/). 



\begin{thebibliography}{10}
\expandafter\ifx\csname url\endcsname\relax
  \def\url#1{\texttt{#1}}\fi
\expandafter\ifx\csname urlprefix\endcsname\relax\def\urlprefix{URL }\fi
\expandafter\ifx\csname href\endcsname\relax
  \def\href#1#2{#2} \def\path#1{#1}\fi

\bibitem{IHTDSC}
L.~Y. Zhang, Y.~Zheng, J.~Weng, C.~Wang, Z.~Shan, K.~Ren, You can access but
  you cannot leak: defending against illegal content redistribution in
  encrypted cloud media center, IEEE Transactions on Dependable and Secure
  Computing (2018).

\bibitem{SteTDSC}
A.~El-Atawy, Q.~Duan, E.~Al-Shaer, A novel class of robust covert channels
  using out-of-order packets, IEEE Transactions on Dependable and Secure
  Computing 14~(2) (2015) 116--129.

\bibitem{JPEG}
G.~K. Wallace, The {JPEG} still picture compression standard, {IEEE}
  transactions on consumer electronics 38~(1) (1992) xviii--xxxiv.

\bibitem{JPEGSteTDSC}
W.~Lu, Y.~Xue, Y.~Yeung, H.~Liu, J.~Huang, Y.~Shi, Secure halftone image
  steganography based on pixel density transition, IEEE Transactions on
  Dependable and Secure Computing (2019).

\bibitem{JPEGSteTDSC2}
Z.~Qian, H.~Zhou, X.~Zhang, W.~Zhang, Separable reversible data hiding in
  encrypted jpeg bitstreams, IEEE Transactions on Dependable and Secure
  Computing 15~(6) (2016) 1055--1067.

\bibitem{STCs}
T.~Filler, J.~Judas, J.~Fridrich, Minimizing additive distortion in
  steganography using syndrome-trellis codes, {IEEE} Transactions on
  Information Forensics and Security. 6~(3-2) (2011) 920--935.

\bibitem{JUNIWARD}
V.~Holub, J.~J. Fridrich, T.~Denemark, Universal distortion function for
  steganography in an arbitrary domain, {EURASIP} Journal of Information
  Security. 2014 (2014) 1.

\bibitem{UERD}
L.~Guo, J.~Ni, W.~Su, C.~Tang, Y.~Shi, Using statistical image model for {JPEG}
  steganography: Uniform embedding revisited, {IEEE} Transactions on
  Information Forensics and Security. 10~(12) (2015) 2669--2680.

\bibitem{RBV}
Q.~Wei, Z.~Yin, Z.~Wang, X.~Zhang, Distortion function based on residual blocks
  for {JPEG} steganography, Multimedia Tools and Applications. 77~(14) (2018)
  17875--17888.

\bibitem{BET}
X.~Hu, J.~Ni, Y.~Shi, Efficient {JPEG} steganography using domain
  transformation of embedding entropy, {IEEE} Signal Processing Letters. 25~(6)
  (2018) 773--777.

\bibitem{GUED}
W.~Su, J.~Ni, X.~Li, Y.~Shi, A new distortion function design for {JPEG}
  steganography using the generalized uniform embedding strategy, {IEEE}
  Transactions on Circuits Systems and Video Technology. 28~(12) (2018)
  3545--3549.

\bibitem{J-MiPOD}
R.~Cogranne, Q.~Giboulot, P.~Bas, Steganography by minimizing statistical
  detectability: The cases of {JPEG} and color images., in: ACM Information
  Hiding and MultiMedia Security (IH\&MMSec), 2020.

\bibitem{Complexity-First-Rule}
B.~Li, S.~Tan, M.~Wang, J.~Huang, Investigation on cost assignment in spatial
  image steganography, {IEEE} Transactions on Information Forensics and
  Security. 9~(8) (2014) 1264--1277.

\bibitem{BBC}
W.~Li, W.~Zhang, K.~Chen, W.~Zhou, N.~Yu, Defining joint distortion for {JPEG}
  steganography, in: R.~B{\"{o}}hme, C.~Pasquini, G.~Boato, P.~Sch{\"{o}}ttle
  (Eds.), Proceedings of the 6th {ACM} Workshop on Information Hiding and
  Multimedia Security, Innsbruck, Austria, June 20-22, 2018, {ACM}, 2018, pp.
  5--16.

\bibitem{BBC++}
Y.~{Wang}, W.~{Li}, W.~{Zhang}, X.~{Yu}, K.~{Liu}, N.~{Yu}, {BBC}++: Enhanced
  block boundary continuity on defining non-additive distortion for jpeg
  steganography, {IEEE} Transactions on Circuits and Systems for Video
  Technology. (2020) 1--1.

\bibitem{BBM}
Y.~{Wang}, W.~{Zhang}, W.~{Li}, N.~{Yu}, Non-additive cost functions for jpeg
  steganography based on block boundary maintenance, {IEEE} Transactions on
  Information Forensics and Security. (2020) 1--1.

\bibitem{DSTC}
C.~Kin-Cleaves, A.~D. Ker, Adaptive steganography in the noisy channel with
  dual-syndrome trellis codes, in: 2018 {IEEE} International Workshop on
  Information Forensics and Security ({WIFS}), IEEE, 2018, pp. 1--7.

\bibitem{SNP}
W.~Sun, J.~Zhou, Y.~Li, M.~Cheung, J.~She, Robust high-capacity watermarking
  over online social network shared images, {IEEE} Transactions on Circuits and
  Systems for Video Technology. (2020).

\bibitem{TCM}
Z.~Zhao, Q.~Guan, H.~Zhang, X.~Zhao, Improving the robustness of adaptive
  steganographic algorithms based on transport channel matching, {IEEE}
  Transactions on Information Forensics and Security. 14~(7) (2019) 1843--1856.

\bibitem{DMAS}
Y.~Zhang, X.~Zhu, C.~Qin, C.~Yang, X.~Luo, Dither modulation based adaptive
  steganography resisting jpeg compression and statistic detection, Multimedia
  Tools and Applications. 77~(14) (2018) 17913--17935.

\bibitem{Robust-steganography-framework}
Y.~Zhang, X.~Luo, C.~Yang, D.~Ye, F.~Liu, A framework of adaptive steganography
  resisting {JPEG} compression and detection, Security and Communication
  Networks 9~(15) (2016) 2957--2971.

\bibitem{GMAS}
X.~Yu, K.~Chen, Y.~Wang, W.~Li, W.~Zhang, N.~Yu, Robust adaptive steganography
  based on generalized dither modulation and expanded embedding domain, Signal
  Processing. 168 (2020).

\bibitem{Downward-robust}
J.~Tao, S.~Li, X.~Zhang, Z.~Wang, Towards robust image steganography, {IEEE}
  Transaction on Circuits and Systems for Video Technology. 29~(2) (2019)
  594--600.

\bibitem{AE-RS}
W.~Lu, J.~Zhang, X.~Zhao, W.~Zhang, J.~Huang, Secure robust jpeg steganography
  based on autoencoder with adaptive bch encoding, IEEE Transactions on
  Circuits and Systems for Video Technology (2020).

\bibitem{CC}
T.~Pevn{\'{y}}, J.~J. Fridrich, Merging markov and {DCT} features for
  multi-class {JPEG} steganalysis, in: E.~J.~D. III, P.~W. Wong (Eds.),
  Security, Steganography, and Watermarking of Multimedia Contents IX, San
  Jose, CA, USA, January 28, 2007, Vol. 6505 of {SPIE} Proceedings, {SPIE},
  2007, p. 650503.

\bibitem{CCPEV}
J.~Kodovsk{\'{y}}, J.~J. Fridrich, Calibration revisited, in: E.~W. Felten,
  J.~Dittmann, J.~J. Fridrich, S.~Craver (Eds.), Multimedia and Security
  Workshop, MM{\&}Sec 2009, Princeton, NJ, USA, September 07 - 08, 2009, {ACM},
  2009, pp. 63--74.

\bibitem{DCTR}
V.~Holub, J.~J. Fridrich, Low-complexity features for {JPEG} steganalysis using
  undecimated {DCT}, {IEEE} Transactions on Information Forensics and Security.
  10~(2) (2015) 219--228.

\bibitem{SRNet}
M.~Boroumand, M.~Chen, J.~Fridrich, Deep residual network for steganalysis of
  digital images, IEEE Transactions on Information Forensics and Security
  14~(5) (2018) 1181--1193.

\bibitem{QIM}
B.~Chen, G.~W. Wornell, Quantization index modulation: {A} class of provably
  good methods for digital watermarking and information embedding, {IEEE}
  Transactions on Information theory 47~(4) (2001) 1423--1443.

\bibitem{QIM-in-DCT}
A.~Miyazaki, A.~Okamoto, Analysis of watermarking systems in the frequency
  domain and its application to design of robust watermarking systems, in:
  {IEEE} International Conference on Acoustics, Speech, and Signal Processing,
  {ICASSP} 2001, 7-11 May, 2001, Salt Palace Convention Center, Salt Lake City,
  Utah, USA, Proceedings, {IEEE}, 2001, pp. 1969--1972.

\bibitem{DBL}
W.~Zhang, X.~Zhang, S.~Wang, A double layered "plus-minus one" data embedding
  scheme, {IEEE} Signal Processing Letters. 14~(11) (2007) 848--851.

\bibitem{Bossbase}
P.~Bas, T.~Filler, T.~Pevn{\'{y}}, "break our steganographic system": The ins
  and outs of organizing {BOSS}, in: T.~Filler, T.~Pevn{\'{y}}, S.~Craver,
  A.~D. Ker (Eds.), Information Hiding - 13th International Conference, {IH}
  2011, Prague, Czech Republic, May 18-20, 2011, Revised Selected Papers, Vol.
  6958 of Lecture Notes in Computer Science, Springer, 2011, pp. 59--70.

\bibitem{JPEG-I}
W.~Luo, J.~Huang, G.~Qiu, {JPEG} error analysis and its applications to digital
  image forensics, IEEE Transactions on Information Forensics and Security
  5~(3) (2010) 480--491.

\bibitem{ensemble}
J.~Kodovsk{\'{y}}, J.~J. Fridrich, V.~Holub, Ensemble classifiers for
  steganalysis of digital media, {IEEE} Transactions on Information Forensics
  and Security. 7~(2) (2012) 432--444.

\bibitem{SI}
W.~Li, K.~Chen, W.~Zhang, H.~Zhou, Y.~Wang, N.~Yu, {JPEG} steganography with
  estimated side-information, IEEE Transactions on Circuits and Systems for
  Video Technology (2019).

\end{thebibliography}

\vskip -2\baselineskip plus -1fil

\begin{IEEEbiography}[{\includegraphics[width=1in,height=1.25in,clip,keepaspectratio]{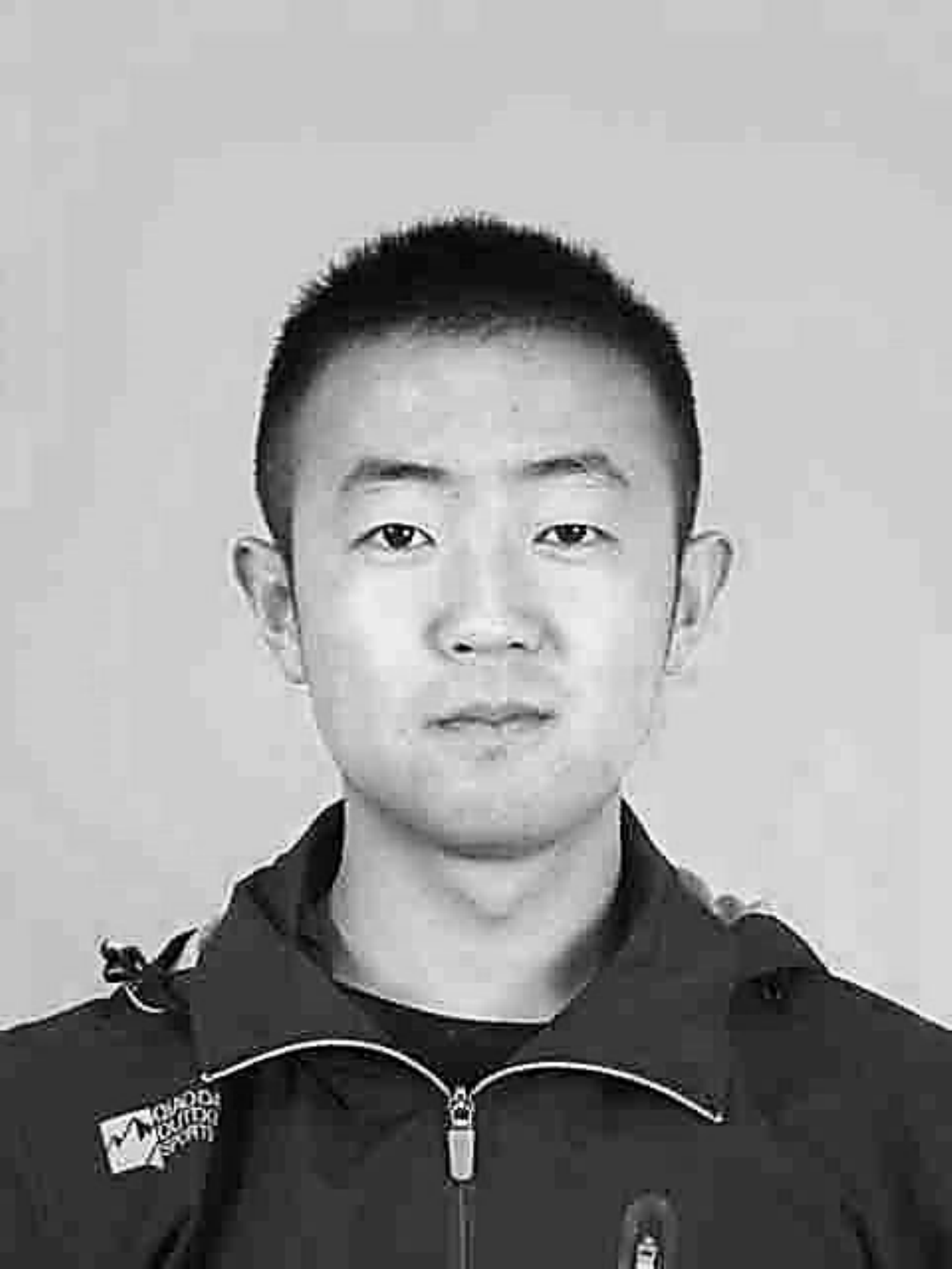}}]{Kai Zeng}
received his B.S. degree in 2019 from the University of Science and Technology of China (USTC). Currently, he is pursuing the
M.S. degree with Key Laboratory of Electromagnetic Space Information, School of Information Science and Technology, University of Science and Technology of China, Hefei. His research interests include information hiding and deep learning.
\end{IEEEbiography}

\begin{IEEEbiography}[{\includegraphics[width=1in,height=1.25in,clip,keepaspectratio]{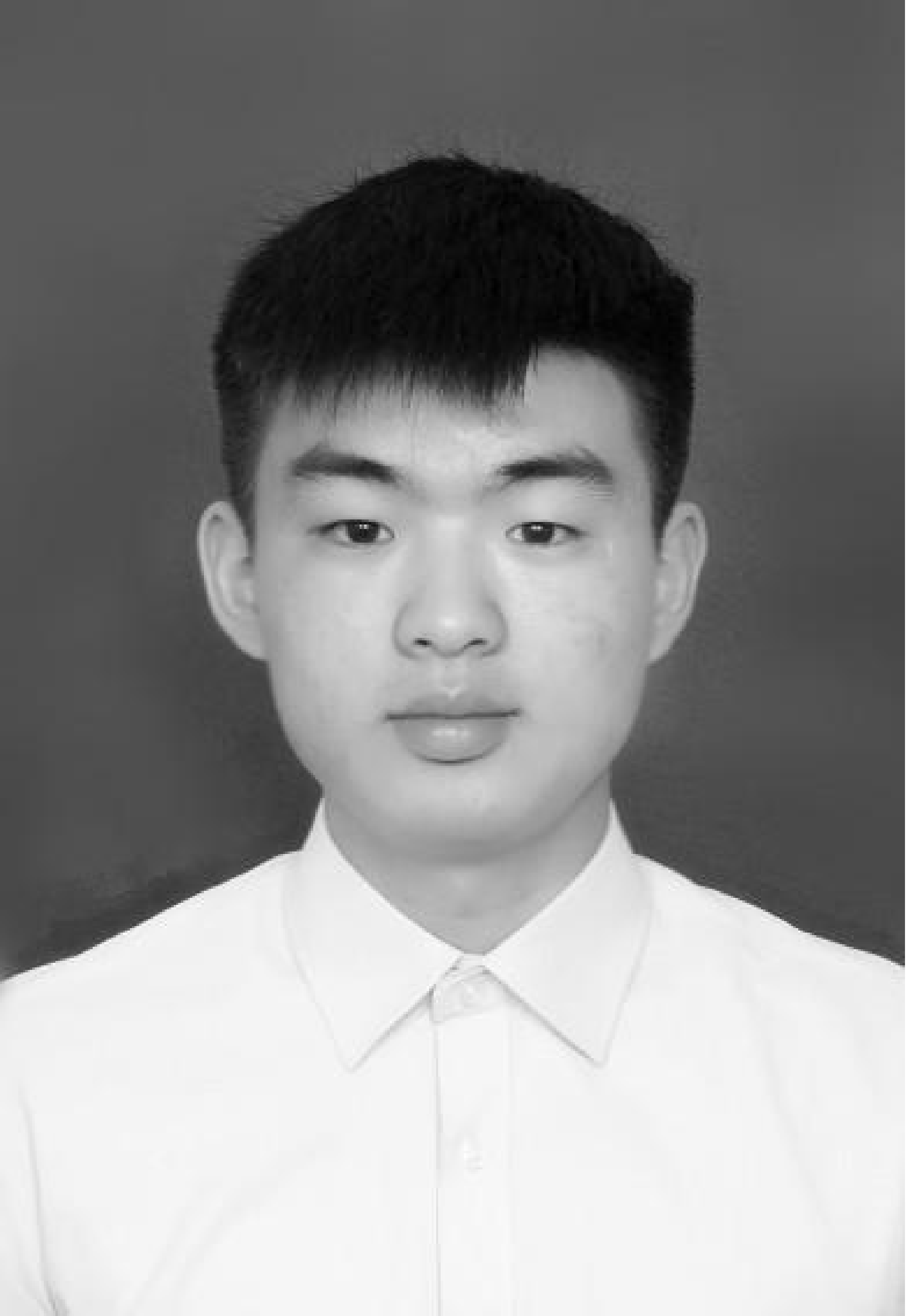}}]{Kejiang Chen}
received his B.S. degree in 2015 from Shanghai University (SHU) and a Ph.D. degree in 2020 from the University of Science and Technology of China (USTC). Currently, he is a postdoctoral researcher at the University of Science and Technology of China. His research interests include information hiding, image processing and deep learning.
\end{IEEEbiography}

\begin{IEEEbiography}[{\includegraphics[width=1in,height=1.25in,clip,keepaspectratio]{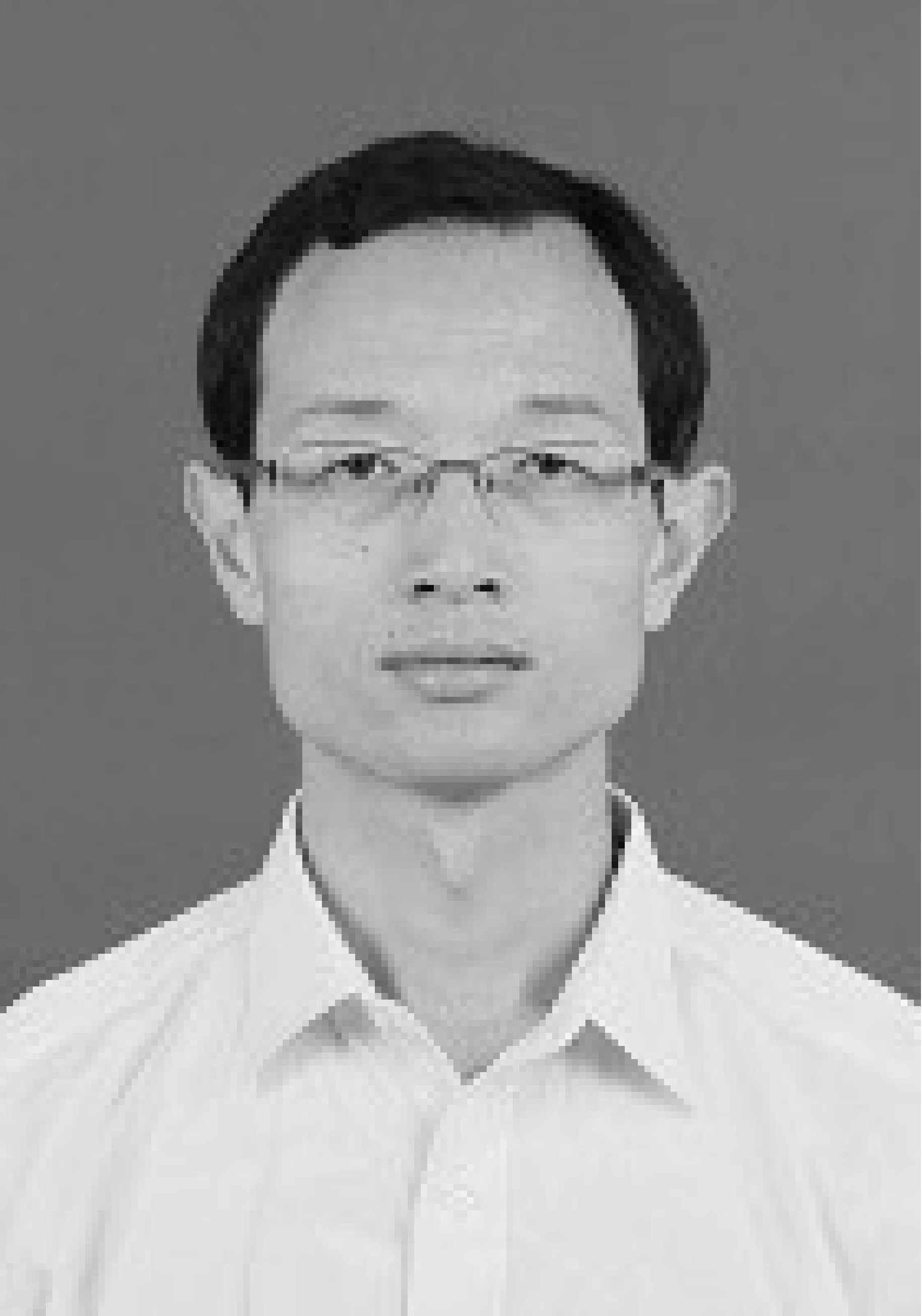}}]{Weiming Zhang}
received his M.S. degree and Ph.D. degree in 2002 and 2005, respectively, from the Zhengzhou Information Science and Technology Institute, P.R. China. Currently, he is a professor with the School of Information Science and Technology, University of Science and Technology of China. His research interests include information hiding and multimedia security.
\end{IEEEbiography}

\begin{IEEEbiography}[{\includegraphics[width=1in,height=1.25in,clip,keepaspectratio]{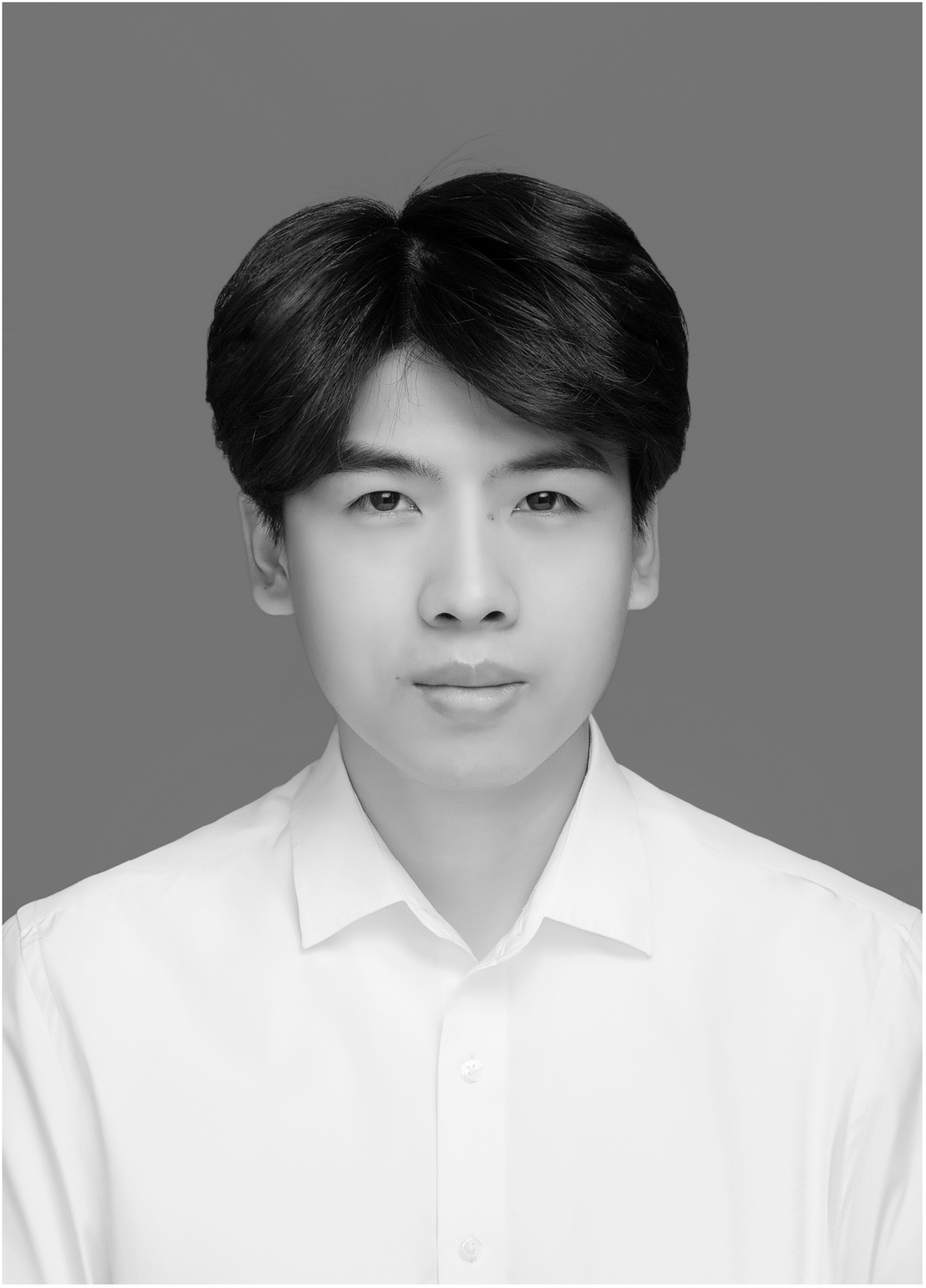}}]{Yaofei Wang}
received the B.S. degree from the School of Physical Science and Technology, Southwest Jiaotong University, in 2017. He is currently pursuing the Ph.D. degree in information security with the University of Science and Technology of China. His research interests include information hiding, image processing, and deep learning.
\end{IEEEbiography}

\begin{IEEEbiography}[{\includegraphics[width=1in,height=1.25in,clip,keepaspectratio]{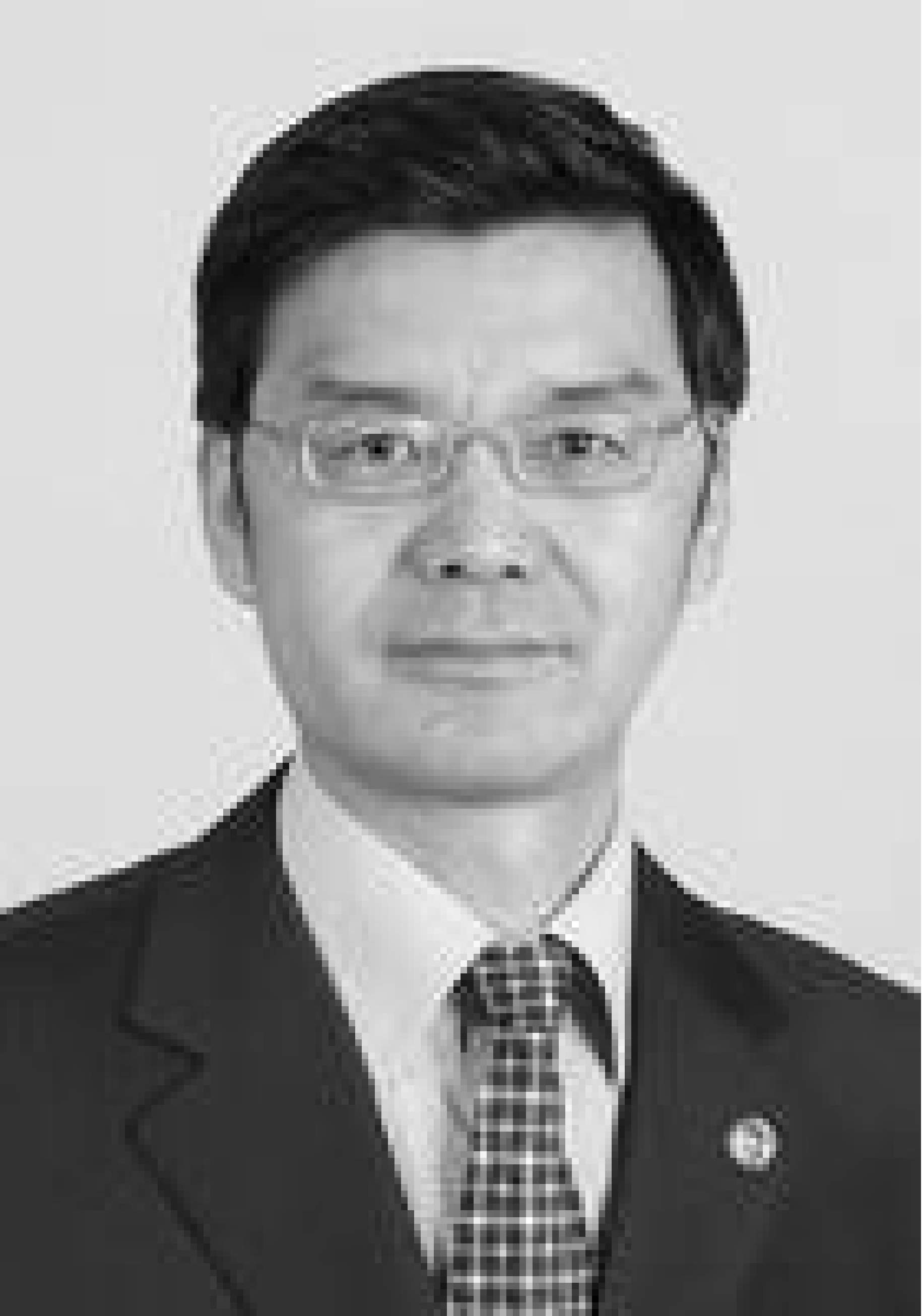}}]{Nenghai Yu}
received his B.S. degree in 1987 from Nanjing University of Posts and Telecommunications, an M.E. degree in 1992 from Tsinghua University and a Ph.D. degree in 2004 from the University of Science and Technology of China, where he is currently a professor. His research interests include multimedia security, multimedia information retrieval, video processing and information hiding.
\end{IEEEbiography}

\end{document}